\documentclass[10pt]{article} 
\usepackage[preprint]{tmlr}

\makeatletter
\newcommand{\anonymousonly}[1]{%
  \if@preprint
  \else
    \if@accepted
    \else
      #1
    \fi
  \fi
}

\newcommand{\publiconly}[1]{%
  \if@preprint
    #1
  \else
    \if@accepted
      #1
    \fi
  \fi
}
\makeatother

\usepackage{amsmath,amsfonts,bm}

\def\eqref#1{equation~\ref{#1}}

\def\1{\bm{1}}

\def\va{{\bm{a}}}

\def\vu{{\bm{u}}}
\def\vv{{\bm{v}}}

\def\vx{{\bm{x}}}
\def\vy{{\bm{y}}}
\def\vz{{\bm{z}}}

\def\mX{{\bm{X}}}

\DeclareMathAlphabet{\mathsfit}{\encodingdefault}{\sfdefault}{m}{sl}
\SetMathAlphabet{\mathsfit}{bold}{\encodingdefault}{\sfdefault}{bx}{n}

\def\sC{{\mathbb{C}}}

\def\sP{{\mathbb{P}}}

\def\sS{{\mathbb{S}}}

\newcommand{\timesend}{\mskip\medmuskip\times}

\usepackage{hyperref}
\usepackage{url}

\usepackage{amsfonts} 
\usepackage{nicefrac}
\usepackage{booktabs} 
\usepackage{graphicx}
\usepackage{siunitx}
\usepackage{amsmath}
\usepackage{tabularx}
\usepackage{makecell}
\usepackage{caption}
\usepackage{subcaption}
\usepackage{xcolor}
\usepackage{longtable}
\usepackage{multirow}

\title{Are You Learning Biological Signal or Shortcuts? Auditing and Mitigating Bias in Protein-Protein Interaction Datasets}

\author{
  \name Judith Bernett \email judith.bernett@fau.de\\
  \addr Department Artificial Intelligence in Biomedical Engineering\\
  Friedrich-Alexander-Universität Erlangen-Nürnberg
  \AND
  \name Anton Spannagl \email anton.spannagl@tum.de \\
  \addr TUM School of Life Sciences\\
  Technical University of Munich
  \AND
  \name Joel Ås \email joelkarl.as@ieo.it \\
  \addr Department of Experimental Oncology\\ 
  IEO European Institute of Oncology IRCCS
   \AND
   \name Markus List \email markus.list@tum.de \\
  TUM School of Life Sciences\\
  Technical University of Munich
   \AND
  \name David B. Blumenthal \email david.b.blumenthal@fau.de \\
  \addr Department Artificial Intelligence in Biomedical Engineering\\
  Friedrich-Alexander-Universität Erlangen-Nürnberg
}

\def\month{08}  
\def\year{2026} 
\def\openreview{\url{https://openreview.net/forum?id=XXXX}} 

\begin{document}

\maketitle

\begin{abstract}
Protein-protein interaction (PPI) databases do not faithfully reflect biological realities. Instead, they are influenced by study and technical biases that distort certain protein and interaction attributes. Machine learning models can exploit these as learning shortcuts if the negative dataset is not constructed with care. So far, the shortcuts introduced during PPI dataset construction have only been examined in isolation.
Here, we systematically characterize both reported and, to our knowledge, previously unreported biases in PPI datasets that lead machine learning models to learn shortcuts instead of biological signal. We analyze HIPPIE, IntAct, and STRING, dedicated PPI databases, as well as two datasets derived from 3D-structural information in the Protein Data Bank (PDB). 
We show that random data splitting introduces strong topological shortcuts. When train-test protein overlap is removed, the resulting datasets still retain usable shortcuts stemming from self-interactions, taxonomic identity, and functional relatedness, whose prevalence interestingly depends on the data source.
We further show that sampling negatives from a set of highly replicated non-interactions, an intuitively appealing choice, can amplify the shortcut stemming from functional relatedness. 
To detect and mitigate these biases, we provide an open Nextflow pipeline that combines similarity-aware, data-loss-minimizing dataset splitting with bias-minimizing negative sampling, both formulated as integer linear programs. 
Its key concept of quantifying biases to minimize them through optimization-based negative sampling can, in principle, be extended to any machine learning problem where the pool of negative candidates is much larger than the positives and is thus of interest also beyond PPI prediction.
\end{abstract}

\section{Introduction}\label{sec:introduction}
Deep machine learning models have a well-documented tendency to learn the ``path of least resistance'', a phenomenon that has spawned an entire field dedicated to studying shortcut learning and its mitigation \citep{geirhos2020shortcut}. 
This has uncovered a series of reported generalization failures that revealed what models were actually paying attention to. Cows were misclassified when displayed on a beach rather than on grass \citep{beery2018recognition}. An object detection model struggled with an elephant randomly inserted into a picture of a man sitting in a room, either missing it entirely or misclassifying it as a chair, depending on its placement \citep{rosenfeld2018elephant}. An X-ray pneumonia detector failed to generalize to images from a new hospital because it had learned to rely on the hospital token in the image and the pneumonia prevalence in that hospital, rather than the lung itself \citep{zech2018variable}. Amazon's HR screening tool was found to systematically favor male candidates \citep{dastin2022amazon}. 
Such cases are cautionary tales for anyone tempted to take strong modeling results at face value, summarized in \citet{geirhos2020shortcut}'s adapted maxim \textit{``Never attribute to high-level abilities that which can be adequately explained by shortcut learning''}\footnote{A blend of Morgan's Canon \textit{``In no case is an animal activity to be interpreted in terms of higher psychological processes if it can be fairly interpreted in terms of processes which stand lower on the scale of psychological evolution and development''} and Hanlon's razor \textit{``Never attribute to malice that which can be adequately explained by stupidity''}}. 
Whether and how effectively a model exploits a shortcut ultimately depends on four components that, together, determine what the model can learn: the model architecture, the training data, the loss function, and the optimization strategy \citep{geirhos2020shortcut}. Before any of these can be adjusted, one must first identify which biases are present in the dataset, a task that is inherently domain-specific and far from trivial.

The issue is pressing in deep learning models for biomedical applications. While we want to be able to harness the full potential of deep learning for healthcare applications, it becomes all the more important to understand what a model is learning, which (latent) features it relies on most, for which subgroups it performs well, and whether it has, indeed, acquired a high-level ability. It is also a domain that is not as easily understood from the human perspective as, e.\,g., object detection or text translation, where a human can more easily verify whether the model performed the task adequately.

In this article, we look at the case of sequence-based protein-protein interaction (PPI) prediction, i.\,e., the task to predict if two proteins bind to each other based on representations of their amino acid sequences. Predicting PPIs is a key task in computational biomedicine, used to understand the functions of understudied proteins, study disease mechanisms, or identify potential drug targets \citep{soleymani2022protein}. The PPI databases, however, contain data from hypothesis-driven experiments, leading to study biases (e.\,g., the overstudy of human disease-related proteins) as well as technical biases (e.\,g., due to low abundance) \citep{schaefer2015correcting,Blumenthal2024-mf}. The negative dataset is usually constructed via random sampling. If the sampling approach is not chosen carefully to mirror certain properties of the positive dataset, unwanted distributional differences can emerge that machine learning models can exploit as shortcuts rather than learning causal biochemical properties, such as matching binding sites. Previously reported properties used as shortcuts included subcellular location \citep{ben2006choosing}, sequence similarity \citep{hamp2015more, bernett2024cracking}, topology \citep{park2012flaws, eid2021systematic, bernett2024cracking, chatterjee2023aibind, zhou2025effectiveness}, taxonomy \citep{hallee2025protein}, the protein language model embeddings \citep{szymborski2026flaw}, and database curation and study bias \citep{szymborski2022rapppid, makrodimitris2020thorough}. Each of these has, however, been studied in isolation, and typically on a single dataset from a single data source. Consequently, it remains unclear which shortcuts dominate, how strongly they interact, whether mitigating one aggravates another, and whether shortcuts differ across PPI databases.

We address this gap with the following contributions: First, we quantify whether certain attributes (self-interactions, taxonomy, topology, sequence similarity, embedding similarity, GO category overlap \citep{aleksander2025go_new}) can be used as a shortcut in commonly used PPI data sources, employing the G-AUDIT bias detection framework \citep{drenkow2025gaudit}. The framework measures whether a given attribute is highly detectable from the input data and whether it has mutual information with the output label. 
Second, we improve \citet{bernett2024cracking}'s previously proposed similarity- and leakage-reducing approach by formulating similarity-aware PPI dataset splitting as an integer linear program (ILP) that additionally minimizes data loss while giving a user control over the resulting split sizes. This formulation is a PPI-specialized adaptation of DataSAIL \citep{joeres2025datasail}.
Third, we also formulate bias-minimizing negative sampling as an ILP. Here, we sample pairs of non-interacting proteins such that the distributions of the shortcut attributes identified in step 1 (self-interaction count, per-protein degree, taxon-pair distribution, and mean GO-BP overlap) are similar in the sets of sampled negatives and input positives, i.\,e., the annotated PPIs. 
We show that datasets derived from 3D protein complex structures curated in the Protein Data Bank (PDB) \citep{berman2000pdb} store a fundamentally different population of interactions than the databases containing data from dedicated interaction-detection assays, limiting their direct use to augment PPI datasets with structural information. 
Further, we show that sampling negatives from a set of highly replicated non-interactions (HRNIs) can amplify existing distributional shifts. 
All analyses are implemented in an open Nextflow pipeline that the community can apply to their own datasets.

\section{Results}

\subsection{Study overview}

\begin{figure}[htb]
    \centering
    \includegraphics[width=\linewidth]{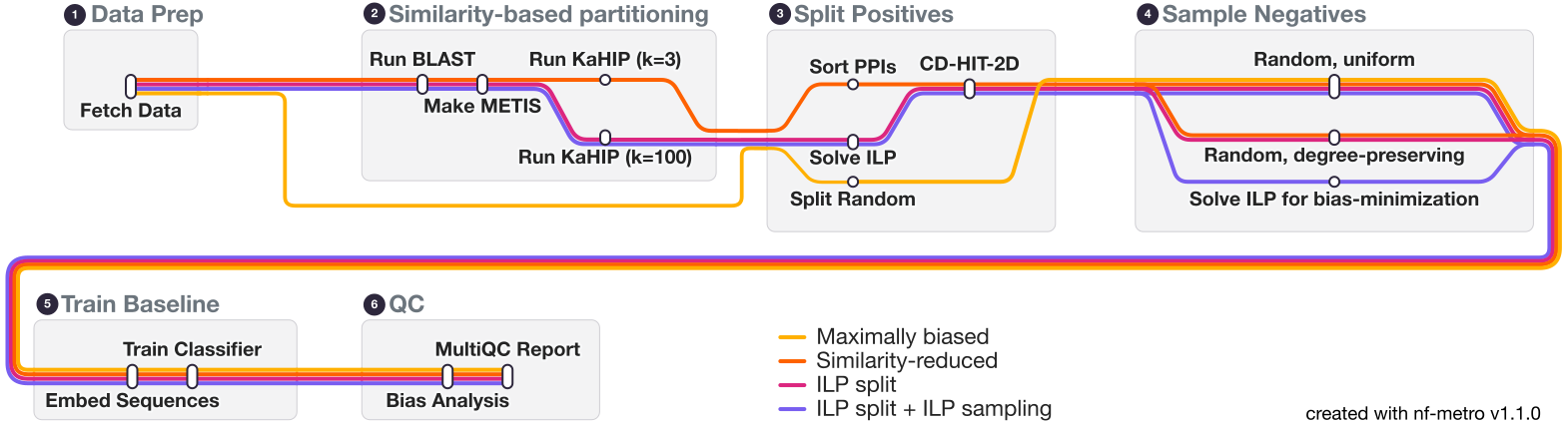}
    \caption{Metro map of the major steps implemented in our Nextflow PPI splitting pipeline for the five configurations outlined in Table \ref{tab:configurations}. If sequence, species, and GO information should be fetched from UniProt, the only required input is a PPI CSV file containing pairs of UniProt IDs. If not, or if they have been pre-fetched, they can additionally be provided to the pipeline. For the ILP split + ILP sampling from HRNI configuration (sampling from a pool of HRNIs), a candidate network is required as additional input. The pipeline can be configured to run in a robustness mode with several random seeds for splitting the positives, sampling the negatives, and training the baseline. Generated with \texttt{seqeralabs/nf-metro}.}
    \label{fig:metro-map}
\end{figure}

All analyses were run with our PPI splitting Nextflow pipeline (Figure \ref{fig:metro-map}). Given a CSV file with PPIs, it fetches sequences, taxonomy, and GO annotations from UniProt \citep{uniprot2025uniprot}, performs similarity-based splitting of the positives into train/validation/test, samples corresponding negatives (if requested, in a bias-reducing manner), trains a baseline classifier, and quantifies the remaining biases in the resulting dataset. Detailed descriptions of all components are provided in Section~\ref{sec:pipeline} in \nameref{sec:methods}. 

The baseline classifier is a Random Forest classifier trained on mean-pooled, concatenated embeddings produced by ESM-2-650M, a widely used BERT-style protein language model trained on huge volumes of amino acid sequences \citep{lin2023esm2}. For all main text results, we fit the Random Forest with 3 different seeds to ensure robustness. It was shown that replacing simpler protein representations with ESM-2 embeddings already improves PPI classifier performance, independent of model architecture \citep{Reim2025-ff}.  For a protein with length $n$, ESM-2-650 produces an $n\times 1280$-dimensional embedding. Mean-pooling over $n$ preserves global sequence properties such as taxonomy, structural class, enzymatic function, and localization \citep{elnaggar2022prott5}, but discards the residue-level information that is crucial to detect binding sites and binding-site compatibility. Therefore, the baseline classifier is expected to perform at chance level on an unbiased dataset. Hence, if the baseline is able to successfully separate positives from negatives, its performance is almost certainly driven by a shortcut rather than by a learned representation of matching binding sites. In addition to ESM-2, we ran all analyses with embeddings produced by ProtT5 \citep{elnaggar2022prott5}, another popular protein language model. Overall, results are similar, as detailed in Appendix~\ref{appendix-prott5}: The utility of the embedding similarity attribute is comparable, the Ridge regressor detects the biases slightly better from the ESM-2 embeddings, and the baseline results are slightly better for ProtT5.

\begin{table}[thb]
\centering
\caption{The five configurations evaluated in this study. Positives are either split randomly, assigned using a similarity-based partitioning of the input proteome (KaHIP, $k=3$), or assigned by minimizing data loss (KaHIP, $k=100$ + ILP). Negatives can be sampled uniformly at random, by preserving positive node degrees, or by minimizing biases using an ILP. The initial candidates for the negatives are either the complement of the input PPI graph or a pre-computed set of HRNIs.}
\label{tab:configurations}
\begin{tabular}{p{4cm}llp{3cm}}
\toprule
\textbf{Configuration} & \textbf{Split of positives} & \textbf{Negative sampling} & \textbf{Negative pool} \\
\midrule
Maximally biased & Random 80/10/10 & Uniform & Complement \\
Similarity-reduced & KaHIP, $k=3$ & Degree-preserving & Complement \\
ILP split & KaHIP, $k=100$ + ILP & Degree-preserving & Complement \\
ILP split + ILP sampling & KaHIP, $k=100$ + ILP & ILP & Complement \\
ILP split + ILP sampling from HRNIs & KaHIP, $k=100$ + ILP & ILP & Highly replicated non-interactions \\
\bottomrule
\end{tabular}
\end{table}

We evaluate the five configurations listed in Table \ref{tab:configurations}, which combine three strategies for splitting the positives with three strategies for sampling the negatives. As both the splitting and sampling procedures contain randomness, they are repeated with three different seeds each. This yields three splits of the positives per configuration and nine datasets, each consisting of training, validation, and test sets with positives and negatives, and hence, 27 Random Forest classifier results per configuration. The splits differ in how strictly they suppress sequence similarity between training and test data and in how much data they discard in doing so. The samplers differ in which statistical properties of the positive set they reproduce in the negative set. The last configuration replaces the pool from which negatives are drawn: instead of sampling from all protein pairs not annotated as interacting, it samples from a precomputed set of highly replicated human non-interactions \citep{as2026negative}, and is therefore only applicable to the human datasets with canonical UniProt sequences.

Each configuration is evaluated on two test sets: a balanced test set and a realistic test set. The balanced test set is sampled using the same strategy as the training and validation sets and contains as many negative samples as positive samples. For the realistic test set, we always sample negatives uniformly and include ten times as many negatives as positives, reflecting that two arbitrary proteins are far more likely not to interact than to interact. The balanced test set measures whether a model can distinguish interacting from non-interacting pairs under the conditions it was trained on, while the realistic test set measures whether it can also do so when the class balance and negative distribution are as expected in practice. Precision, F1, MCC, and AUPRC are always worse on the realistic test set than on the balanced one, as these metrics are sensitive to class imbalance: a tenfold increase in negatives will turn the same false-positive rate into many more false positives.

We ran all of these analyses on nine datasets, spanning experimentally curated interactions (HIPPIE, IntAct), score-channel subsets of STRING, and two structural datasets derived from the PDB (PINDER, PDB-Dimers). Table \ref{tab:ppi-datasets} summarizes their sizes and characteristics, detailed dataset descriptions are provided in Section \ref{sec:datasets} in \nameref{sec:methods}.

\subsection{Performance under the maximally biased setting}

\begin{figure}[!htb]
    \centering
    \includegraphics[width=\linewidth]{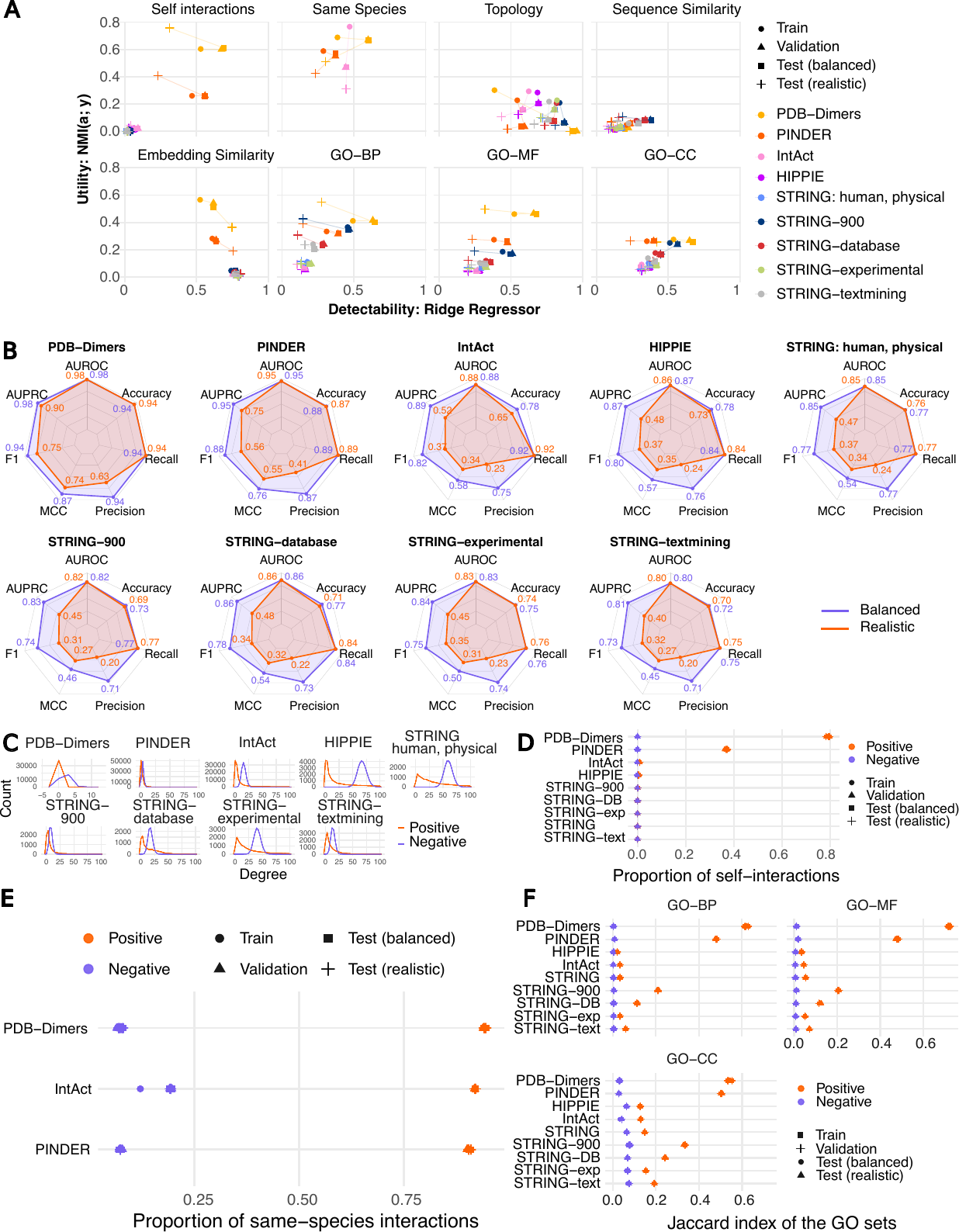}
    \caption{Overview of the maximally biased setting. \emph{(caption continued on next page)}}
    \label{fig:max-biased}
\end{figure}
\addtocounter{figure}{-1}
\begin{figure} [t!]
  \caption{\textbf{A} Bias analysis per attribute $\va$ for the split/sampling seed combination 1234/42. The higher the utility (NMI between $\va$ and the label $\vy$) and detectability (Spearman $\rho$ between $\va$ and $\va_{pred}$, predicted by a Ridge regressor from the input), the higher the shortcut potential for $\va$. \textbf{B} Mean performance of the baseline across 27 seed combinations (3 for splitting, 3 for sampling, 3 for training), all standard deviations (SDs) are $\leq 0.01$ (see Table \ref{tab:all_baseline_results}). The realistic test set has 10 times as many negatives as positives. 
  \textbf{C} Degree distribution of each protein in the positive (orange) vs. the negative (purple) train set (one line per split/sampling seed). Degrees cropped at 100 for readability.
  \textbf{D} Ratio of self-interactions/all interactions. One point per split/sampling seed.
  \textbf{E} Ratio of same-species interactions/all interactions. One point per split/sampling seed.
  \textbf{F} Mean Jaccard indices of the GO terms annotated for the interaction candidates. One point per split/sampling seed. \emph{(continued)}}
\end{figure}

As shown in Figure \ref{fig:max-biased}B and Supplementary Table \ref{tab:all_baseline_results}, a simple Random Forest baseline achieves strong performance in the maximally biased setting (AUROC 0.80-0.98 on the balanced test set), exploiting different shortcuts depending on the dataset. 
To identify which ones, we employ the G-AUDIT \citep{drenkow2025gaudit} bias analysis (Figure \ref{fig:max-biased}A), in which the strongest shortcut candidates are attributes that are both highly detectable from the input and highly informative about the label. We run this bias analysis on only one seed combination since the standard deviations between all 27 Random Forest classifier runs were $\leq 0.01$ and the properties in question do not vary between the different sets (Figure \ref{fig:max-biased}C-F). Embedding similarity is highly detectable by construction, since the embeddings are the input to the detector, so for this attribute, we are only interested in the normalized mutual information (NMI), i.\,e., in whether the embedding implicitly encodes important properties of interaction. This does not seem to be the case, as the NMI on the structural datasets is driven by the self-interactions (see below).

Topology is the attribute with the strongest detectability across all datasets and, for most of them, also the highest utility. For a pair of proteins, topology is defined as their average degree ratio, where a protein's degree ratio is its positive degree in the training set divided by its total training degree \citep{chatterjee2023aibind}. 
As Figure \ref{fig:max-biased}C shows, positive and negative pairs differ strongly in degree for IntAct, HIPPIE, STRING: human/physical, and STRING-experimental, making them largely separable by degree alone. 
These are also the datasets where topology retains high utility across all splits except the realistic test set, which likely drives the baseline's strong performance on them.
The shortcut is only available in the maximally biased setting, because in all other configurations, there is no train-test protein overlap, so it cannot be computed. 
These results illustrate the danger of train-test overlap: it allows the model to ignore the input entirely and still perform well based on protein identity alone \citep{park2012flaws, eid2021systematic, bernett2024cracking, chatterjee2023aibind, zhou2025effectiveness}.

For the two structural datasets, PDB-Dimers and PINDER, the self-interaction shortcut dominates.
Figure \ref{fig:max-biased}D shows that 80\% of all positive interactions in PDB-Dimers are self-interactions, and roughly 40\% in PINDER. 
Because negatives are sampled at random and therefore rarely pair a protein with itself, whether two input proteins are identical becomes both highly detectable and highly predictive of the label (hence, the strong baseline performance in Figure \ref{fig:max-biased}B). 
The shortcut also contaminates three of the other attributes: a self-pair is, by construction, same-species, has an embedding cosine similarity of 1, and has a GO Jaccard index of 1. Sequence similarity is unaffected, because we exclude self-interactions for its computation to be able to separate the similarity signal from the identity signal.
Self-interactions are a real shortcut rather than an attribute we want to learn because in reality: Many proteins never bind copies of themselves, and most interactions occur between different proteins. Apart from that, the distributional differences between the PDB and other databases pose a real issue for using the PDB to augment PPI datasets with structural information. The PDB apparently annotates a very different population of interactions, weighted heavily toward homo(di)mers rather than hetero(di)mers. This is unfortunate, since with AlphaFoldDB \citep{varadi2024alphafolddb} providing structures for individual proteins and the PDB providing structures for complexes, structure-informed PPI prediction should be within reach. 

Sequence similarity, in contrast, turns out to be a weaker shortcut than prior work would suggest \citep{hamp2015more, bernett2024cracking}. Supplementary Figure \ref{fig:similarity} shows the distribution of the percentage of identical residues between two positive versus two negative pairs. The negatives have a strong peak at zero across all datasets, as expected. The largest distributional difference can be seen for STRING-900, STRING-database, and STRING-textmining, which is where the attribute also reaches its highest shortcut potential in Panel A. 
A reason could be that the STRING-database and STRING-textmining subsets might contain more paralogs. Gene duplication produces sequence-similar proteins that also tend to share pathway and complex membership \citep{pereira2007evolution}. STRING's use of homology transfer between organisms and the name-based nature of text mining might amplify this further \citep{szklarczyk2025string}. 
While the utility of sequence similarity stays low throughout, suppressing it between splits nevertheless remains necessary, because isoforms, modified, or mutated sequences, paralogs, or closely related homologs should not be placed in different splits if we want to assess whether our model has learned biological mechanisms of binding. This matters most for the structural datasets, where the same protein occurs in many slightly different variants.

The taxonomy bias previously described by \citet{hallee2025protein} is relevant for the three multi-species datasets IntAct, PDB-Dimers, and PINDER (Figure \ref{fig:max-biased}A, E). About 90\% of their annotated PPIs occur within the same species, whereas fewer than 20\% of the randomly sampled negative pairs do, resulting in a high measured utility. 
Figure \ref{fig:max-biased}A further shows that species can be inferred reasonably well from the embeddings (mean detectability 0.41). 
Inter-species PPIs are, however, biologically possible (e.\,g., host-virus interactions), and we would ideally want our model to detect whether two proteins bind based on their sequence or structure, independent of taxonomy. 
The taxonomy attribute is therefore an unwanted shortcut that must be corrected for, as failing to do so distorts the resulting performance estimates (see the IntAct results in Panel B). 

Finally, among the non-structural datasets, the GO category overlap is especially predictive of STRING-900, STRING-database, and STRING-textmining, and is also sufficiently detectable from the embeddings to be exploitable. 
Figure \ref{fig:max-biased}F shows that PPIs in these three subsets are more functionally related than those in the purely experimental sources (HIPPIE, IntAct, and STRING-experimental) and in STRING: human/physical. 
This observation makes sense since the STRING database channel uses pathway and complex annotations (which are functionally related by design) and the textmining channel identifies protein co-mentions in the literature, which are also more likely to reflect shared biological function or pathway membership \citep{szklarczyk2025string}. 
Since STRING-900 aggregates all evidence channels, it inherits the functional enrichment from these two. 
This is an important pitfall to keep in mind: intuitively, a user might select only the most confident STRING interactions to train their model, expecting higher-quality labels to yield more trustworthy results. Instead, part of the signal in this subset stems from functional relatedness encoded in the embeddings, which is not as pronounced in a purely experimental dataset. Hence, reported performances might be inflated, as they would not reflect the model's true ability to detect genuine physical interactions in an experimental setting. 

\subsection{Similarity-reduced versus ILP-based splitting}

\begin{figure}[htb]
    \centering
    \includegraphics[width=0.94\linewidth]{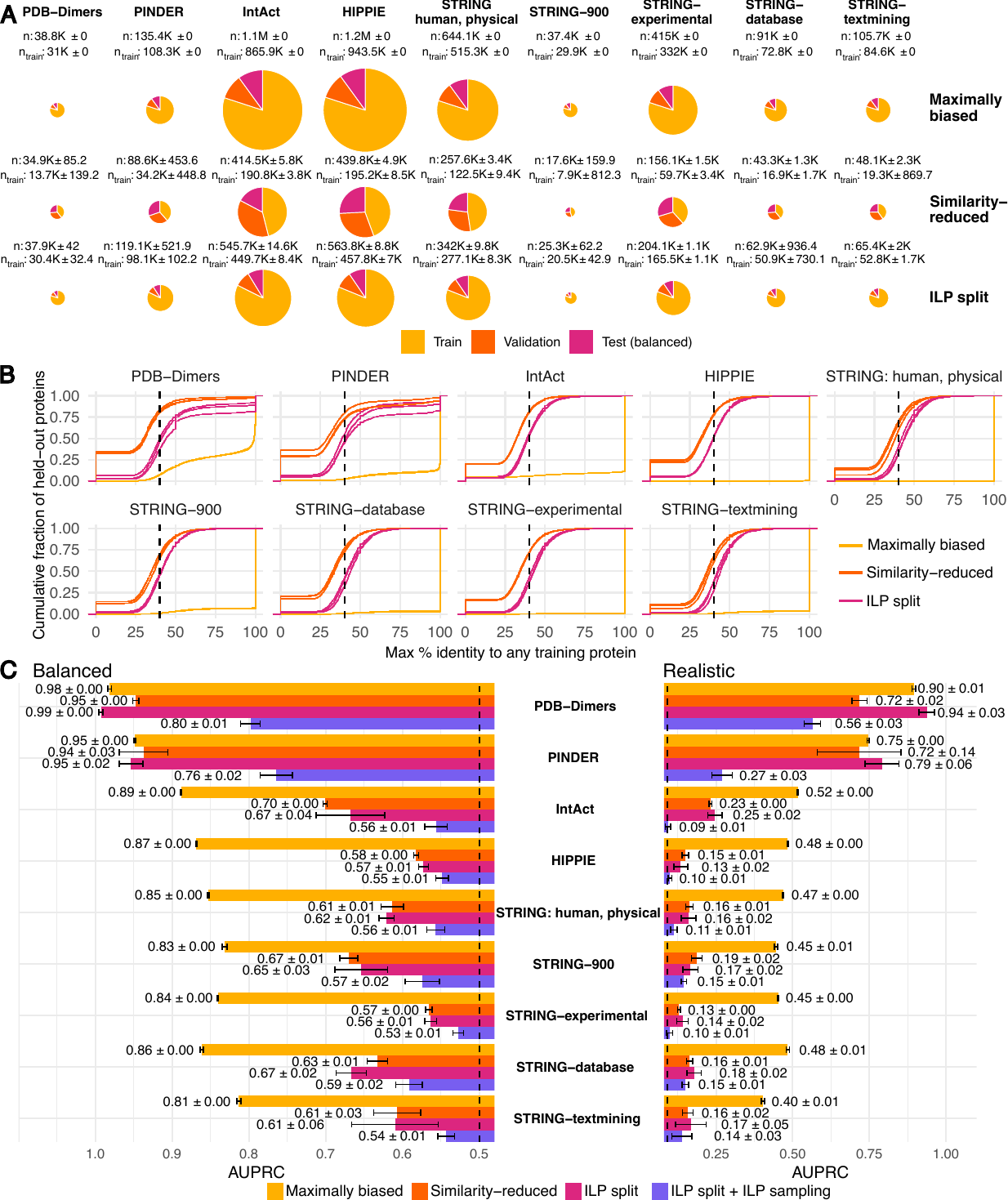}
    \caption{Comparison of the similarity-reduced and the ILP-based settings. \textbf{A} Dataset sizes for the positives (mean ± standard deviation (SD) over 3 split seeds). \textbf{B} Maximum percentage identity (BLAST) of test proteins to proteins in the respective training set, one line/split seed. The dotted line indicates 40\% identity, the threshold for CD-HIT-2D. \textbf{C}  Mean ± SD performance of the Random Forest over 27 seed combinations (3/splitting, 3/sampling, 3/training). The dotted line indicates the chance performance ($0.5$ and $1/11$).}
    \label{fig:blackbox_vs_ilp}
\end{figure}

The similarity-reduced split previously proposed by \citet{bernett2024cracking} has two practical drawbacks: It offers no control over the sizes of the resulting training, validation, and test sets, and, because it partitions the entire proteome before assigning PPIs to splits, a large number of interactions are discarded because their endpoints are part of different blocks. We therefore adapt the DataSAIL \citep{joeres2025datasail} approach by formulating splitting as an ILP that minimizes the data loss while enforcing specified train/validation/test fractions.
Figure \ref{fig:blackbox_vs_ilp}A shows how the ILP-based split fulfills the specified 80/10/10 split while retaining substantially more interactions for training. For PDB-Dimers and PINDER specifically, overall data loss always remains low due to the self-interactions. Because a self-interaction has both endpoints in the same cluster, it can never be split apart.

Figure \ref{fig:blackbox_vs_ilp}B shows that both the similarity-reducing and ILP-based approaches successfully reduce sequence similarity between the training and test sets. The similarity-reducing approach produces more dissimilar splits than the ILP approach because at $k=3$, KaFFPa directly minimizes the edge cut among the three groups, which later form the splits. With $k=100$, similarity is minimized between 100 clusters, and the ILP then regroups them into three splits according to data loss alone, so a test cluster can still be rather similar to a train cluster. To limit this, we apply CD-HIT-2D to remove any validation or test PPI whose interactors have more than 40\% identity to a training sequence, indicated by the dotted line (see Supplementary Figure \ref{fig:cdhit} for the proportion of discarded PPIs). 

Figure \ref{fig:blackbox_vs_ilp}C and Supplementary Table \ref{tab:all_baseline_results} shows the effect on the Random Forest baseline. Across all datasets and configurations, changing the splitting seed resulted in the largest performance changes. The goal is to obtain chance-level performance, i.\,e., an AUPRC of 0.5 on the balanced set and roughly 0.09 on the realistic test set. 
The balanced test set checks in-distribution whether training succeeded, i.\,e., whether the model predicts PPIs without relying on the shortcuts removed by sampling. The realistic test set simulates the application scenario, in which negatives are more abundant and follow no controlled distribution. 
HIPPIE and STRING-experimental come closest, falling from 0.87±00 and 0.84±00 balanced AUPRC in the maximally biased setting to between 0.55 and 0.59 in both settings. The AUPRCs of IntAct, STRING: human/physical, STRING-900, STRING-database, and STRING-textmining remain slightly inflated at mean performances of 0.61--0.68 (before, 0.81--0.89). 
The performance on the structural datasets PDB-Dimers and PINDER remains around 0.95 due to the self-interaction shortcut, which cannot be removed by the splitting strategy. The splitting strategy also cannot remove the taxonomy shortcut in IntAct, and the GO overlap shortcut in the STRING datasets. All three come from distributional differences between the positives and the negatives, so addressing them requires intervening on the negatives.

\subsection{Controlled bias-reduction through the ILP-based negative sampling}

\begin{figure}[htb]
    \centering
    \includegraphics[width=\linewidth]{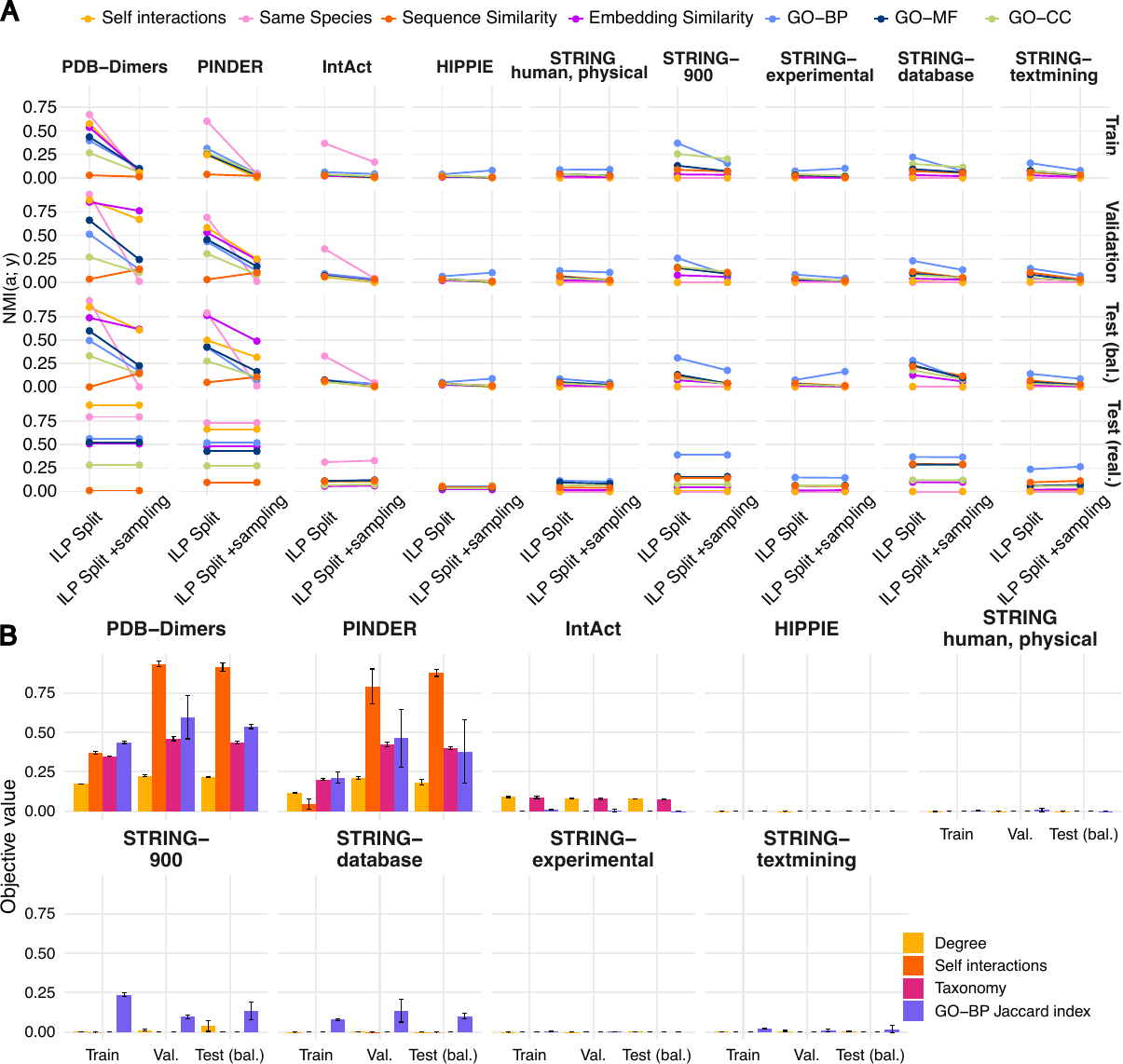}
    \caption{Influence of ILP-based negative sampling. \textbf{A} Decrease in normalized mutual information from the ILP-based split (+degree-preserving random sampling) to the ILP-based split + ILP-based sampling  for the split/sampling seed combination 1234/42. \textbf{B} Composition of the remaining mean $\pm$ SD objective value of the ILP-based negative sampling across all nine (3 splitting, 3 sampling) seed combinations. The same species term was only included for PDB-Dimers, PINDER, and IntAct.}
    \label{fig:ilp_sampling}
\end{figure}

To further mitigate the biases we identified through the G-AUDIT analyses (degree shortcut, GO overlap shortcut, self-interaction shortcut, and taxonomy shortcut), we developed a bias-minimizing ILP-based negative sampler that jointly minimizes the underlying distributional differences between sampled negatives and positives (ILP split + ILP sampling configuration). The ILP proves optimality for every dataset and split for the sampled candidate pool, so any residual bias stems from properties of the negative pool. Results for other pool sizes are displayed in Figures \ref{fig:pool_size_obj} and \ref{fig:pool_go_rf}. Overall, this configuration yields the least-biased datasets. Compared to the ILP split configuration, most attributes now carry less mutual information (Figure~\ref{fig:ilp_sampling}A).
Figure~\ref{fig:ilp_sampling}B shows the composition of the remaining objective value, which sums up the bias terms for self-interactions, taxonomy, degree, and mean GO-BP Jaccard overlap. PDB-Dimers and PINDER reach much higher residual objectives than the other datasets. 

Importantly, the terms are not independent. For the structural datasets PDB-Dimer and PINDER, the self-interaction deviation is the largest remaining bias term that cannot be minimized further, because a negative self-interaction can only be sampled for a protein without a positive one. Both datasets, however, contain positive self-interactions for the majority of the covered proteins, making it impossible for the ILP-based negative sampler to match the numbers of positive and negative self-interactions. 
The problem is made worse by the split, since self-interactions are never discarded, which raises their share in the smaller splits. 
In PDB-Dimers, they make up more than 90\% of validation and test interactions; in PINDER, this imbalance is only that severe in the test set (88\%), which is correspondingly where its self-interaction bias is the highest. The failure to satisfy the self-interaction term propagates to the GO-BP term since any negative sampled for a protein with only self-interactions will contribute a lower Jaccard index to the negative mean.

For the taxonomy term, which matches the number of interactions per taxon pair, the residual in PDB-Dimers and PINDER is driven mainly by human-human interactions. Human proteins account for roughly a quarter of all proteins in both datasets, so negatives are disproportionately likely to be drawn from human proteins (up to $3.7$ times more negative human-human interactions). The remaining bias is distributed across rare taxon combinations, which the logarithmic weighting penalizes more heavily.
In IntAct, human-human pairs account for the vast majority of positives, so a small relative mismatch (around 1.25–1.27 times the target) leads to a nontrivial absolute contribution.

For STRING-900 and STRING-database, GO-BP overlap is the largest remaining bias term. Figure~\ref{fig:ilp_sampling}A shows that the mutual information between the GO categories and the label is reduced but not eliminated, most likely because functionally similar pairs are more likely to already be annotated as interacting in STRING-database, leaving few functionally related candidates available as negatives. 
For HIPPIE, STRING: human/physical, STRING-experimental, and STRING-textmining, all four terms are reduced to near zero, though their bias was low to begin with.

Returning to the baseline results in Figure~\ref{fig:blackbox_vs_ilp}C and Supplementary Table \ref{tab:all_baseline_results}, ILP-based sampling yields the lowest performance on every dataset. For all datasets except PDB-Dimers and PINDER, balanced AUPRC now falls between $0.51$ and $0.62$, and realistic AUPRC between $0.07$ and $0.23$. Excluding the GO-BP bias term re-inflates the STRING datasets but does not impact HIPPIE or IntAct (Figure \ref{fig:pool_go_rf}).

\subsection{The influence of highly replicated negatives}

\begin{figure}[htb]
    \centering
    \includegraphics[width=\linewidth]{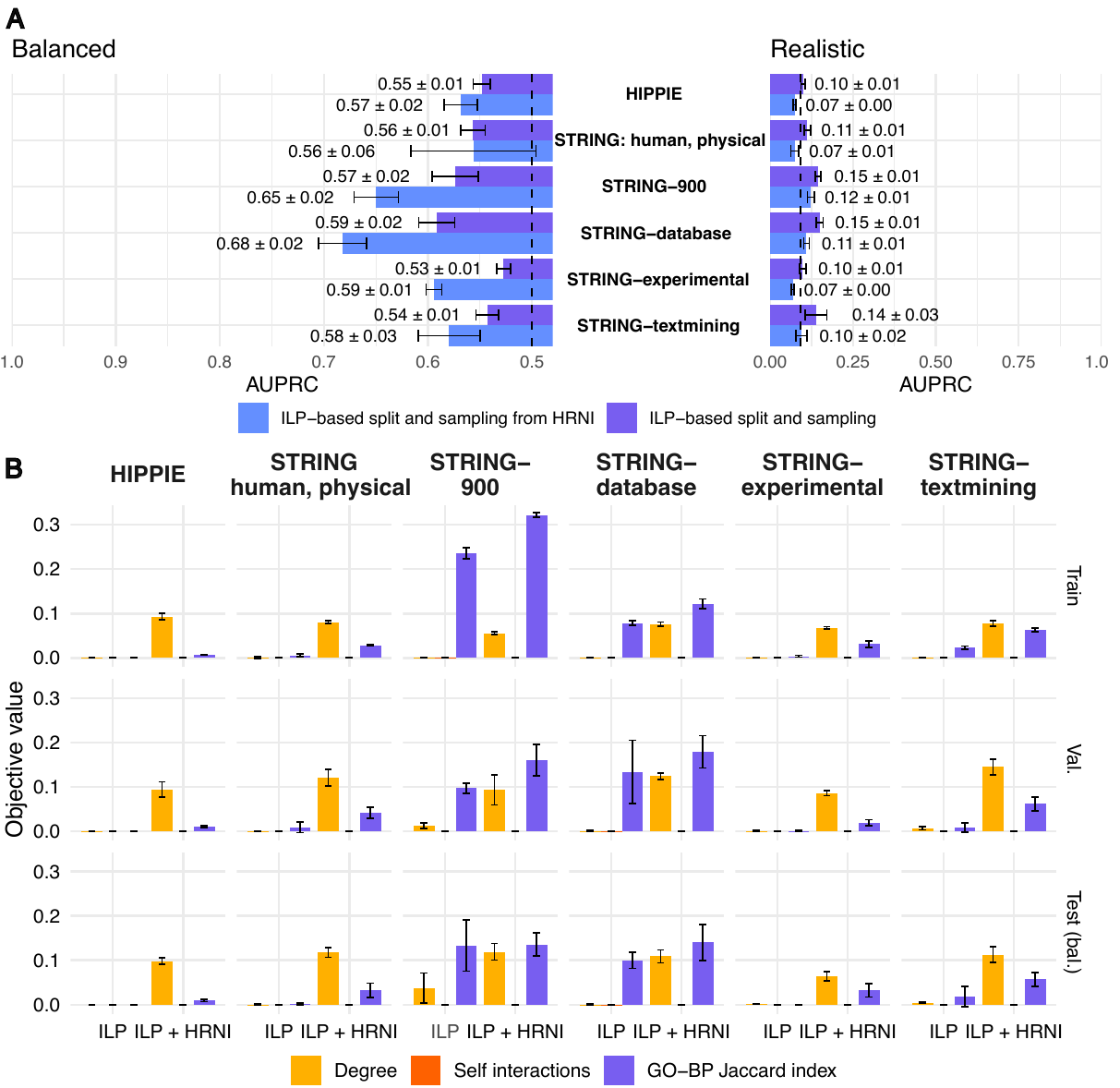}
    \caption{Influence of using a pre-defined set of highly replicated non-interactions for sampling with the ILP. \textbf{A} Mean ± standard deviation of the performance of the Random Forest over 27 seed combinations (3/splitting, 3/sampling, 3/training). \textbf{B} Composition of the remaining mean ± SD objective values of the ILP-based negative sampling runs across all nine (3 splitting, 3 sampling) seed combinations.}
    \label{fig:hcn}
\end{figure}

All configurations so far draw negatives from the complement of the PPI graph, which conflates untested pairs with genuine non-interactors. We assess the consequences of sampling negatives from a precomputed set of HRNIs by constructing them from IntAct pairs of human proteins that were tested together at least 3 times but were never reported to interact \citep{as2026negative}.
Figure \ref{fig:hcn}A shows that performance on the balanced test set increases substantially for all STRING subsets, robustly across all tested seeds. Figure \ref{fig:prott5_rf} and Table \ref{tab:prott5_performance} show the same trends for the ProtT5 embeddings with at most a 3\% difference (generally, AUPRCs achieved with ProtT5 are around 1\% higher than those achieved with ESM-2). This is unlikely the result of higher-quality negatives improving the baseline's ability to learn interaction features, since performance on the realistic dataset remains at chance level. Instead, it points to a distributional difference between the HRNI and the positives that cannot be sampled away by choosing a suitable subset.

The bias terms of the residual objective value (Figure \ref{fig:hcn}B) support this interpretation, as the residual objective term increases across all datasets. The degree term rises for all of them, consistent with the HRNI set spanning only $\num{13557}$ unique proteins (Table \ref{tab:ppi-datasets}), which is too few to satisfy the degree constraint for many proteins. The self-interaction term is negligible throughout, as expected.

The most substantial issue is in the GO-BP term, which appears for HIPPIE, STRING: human/physical, and STRING-experimental, where it was previously near zero, and worsens for the remaining subsets. 
This makes biological sense, since protein pairs confirmed not to interact are less likely to share biological processes, making it infeasible to sample the GO-BP bias away. Indeed, only around $5\%$ of the 25 million HRNI pairs have a GO-BP Jaccard index greater than 0, and for only $1.5\%$ is it greater than 0.1. For this reason, dropping the GO-BP term also has no notable impact on the baseline results in this setting (Figure \ref{fig:pool_go_rf}). HIPPIE is not strongly affected because its positives are not strongly functionally enriched to begin with.
When drawing negatives from a high-confidence set, one should therefore consider which distributional shifts this introduces and whether they are desirable. 

\section{Discussion}

We show that the splitting strategy and the negative sampling approach substantially affect which shortcuts a PPI prediction model can exploit, and that the magnitude of these shortcuts varies across databases. A Random Forest baseline trained on protein embeddings can reach near-perfect performance under a naive random split. While removing the train-test protein overlap and minimizing inter-split sequence similarity are effective countermeasures, several datasets retain strong shortcuts because of distributional differences between the positive and negative sets: for the PDB-derived datasets, the high number of self-interactions is the largest issue. For multi-species datasets, intra-taxon pairs are less likely to be sampled, but taxonomy can be inferred well from embeddings. Finally, the STRING subsets that use database annotations and text mining annotate pairs that are more functionally enriched than expected by chance, which can also be detected sufficiently well from the embeddings. We jointly align these distributions using an ILP-based negative sampler. 
It decreases the baseline to chance level across all datasets except PDB-Dimers and PINDER, where the self-interaction bias cannot be minimized any further, since most proteins in these datasets have positive self-interactions and a negative self-interaction can only be drawn for a protein that does not already have a positive one. Chance-level performance is the expected outcome, since a low-complexity model like our Random Forest baseline without a mechanism to represent an interface should not be able to separate the classes. For STRING-900 and STRING-database, a residual functional-relatedness bias remains because too few functionally related negatives are available to match the positives. 

Our results have practical implications for anyone building or evaluating a PPI prediction model. First, reported performance figures are only interpretable together with the splitting and sampling strategy that produced them. A high AUPRC on a randomly split, uniformly sampled dataset says little about generalization to unseen proteins. 
Second, the choice of the PPI database matters independently of the modeling approach. The PDB-derived datasets encode a fundamentally different population of interactions, weighted heavily toward homo(di)mers, which limits their use for augmenting PPI datasets with structural information. 
This imbalance present in the PDB may affect complex structure prediction as well: While \citet{han2026nvidia} reported high-confidence AlphaFold-Multimer \citep{evans2021protein} predictions for roughly 7\% of 23M predicted homodimers, only 0.7\% of 7.6M predicted heterodimers were high-confidence. Furthermore, these high-confidence heterodimers were enriched for homodimer-like properties such as similar chain length and high inter-chain sequence identity. Third, drawing negatives from a high-confidence pool, while intuitively appealing, can introduce or worsen distributional shifts, particularly for the STRING datasets, since highly replicated negatives tend not to be functionally related. For HIPPIE, which does not suffer from this distributional bias, baseline results stayed approximately the same. It is, therefore, possible that a model with the capabilities to grasp the biochemical mechanisms of interactions can benefit from the less noisy signal of the highly replicated negatives. 

Our analyses have several technical limitations: First, the pipeline only supports the ESM-2 and ProtT5 embedding models. Different embedding models may encode taxonomy, degree, or functional information to varying extents, which can shift detectability and potentially reorder which shortcuts dominate for a given dataset. 
Second, our attribute set is not exhaustive, and additional biases, such as protein length or experimental detection method, may be present; however, more terms can be added to the bias analysis and the sampler. 
Third, we quantify detectability using a Ridge regressor. Replacing it with a GPU-based estimator would reduce the runtime and memory consumption (see Appendix \ref{appendix-runtime}), while providing a tighter lower bound on detectability than a simple linear model.
Fourth, the sampling ILP is solved over a stratified candidate pool of four times the positive count rather than over all admissible negatives, because solving the ILP over the full pool is computationally prohibitive. The returned solution is optimal for the sampled pool, but a better-matched negative set may exist. We assessed the influence of scaling the pool down ($2\timesend$ the positive count) and up ($8\timesend$ the positive count). Generally, larger pools led to smaller residual objective values (Figure \ref{fig:pool_size_obj}) and less biased datasets (i.\,e., closer to chance-level baseline performance; see Figure \ref{fig:pool_go_rf}), but the change from $4\timesend$ to $8\timesend$ is less pronounced than from $2\timesend$ to $4\timesend$.
Fifth, the high-confidence negatives were derived only from human IntAct pairs. The same construction could, in principle, be extended across all of IntAct, but it does not apply to the PDB-based datasets, whose entries often contain modified rather than canonical sequences, so we cannot assume that a modified sequence behaves like its canonical counterpart with respect to interactions. 
Sixth, even our most rigorous mitigation strategy, ILP-based sampling, leaves residual bias in PDB-Dimers and PINDER, driven by their high proportion of self-interactions. Addressing this fully would likely require complementary strategies such as positive downsampling or loss reweighting, which we did not evaluate empirically.

It is also important to note that the functional relatedness (GO overlap) bias attribute we identified via the G-AUDIT analysis is \emph{not only} a confounder: proteins in the same pathway, cellular component, or complex genuinely do interact more often, so matching the GO overlap of the negatives to that of the positives removes real signal along with the artifact. Nonetheless, functional relatedness is \emph{also} a confounder because it holds at the level of pathways and complexes rather than individual pairs, so it cannot distinguish which members of a complex actually bind, which is precisely what a PPI model should predict. For this work, we therefore decided to treat GO overlap as a bias attribute and include it in the objective of our ILP-based negative sampler. Setting the GO term to zero re-inflated the STRING datasets but did not impact HIPPIE or IntAct (Figure \ref{fig:pool_go_rf}).

We release the splitting and bias analysis pipeline as an open tool so that the community can apply the same bias-aware procedures to their own data. It can be adapted to other pair-input prediction tasks, such as domain-domain, protein-ligand, or gene-gene interactions, by substituting the similarity definition and the bias terms. Moreover, specific bias terms can be switched off for negative sampling, e.\,g., users who decide to treat GO overlap as signal can simply set the weight of the corresponding component from the ILP objective of the negative sampler to zero. We have investigated the effects of this in Figure \ref{fig:pool_go_rf} and, as expected, the results for the balanced STRING test sets, which suffer from the GO-BP shortcut, become inflated, while the realistic test set results stay random. For the other datasets, no performance difference on both the balanced and the realistic test set can be observed.

Our framework could also be combined with the bias auditing steps proposed by \citet{eid2021systematic}, who audit a model by randomizing or masking input features or model components, testing generalization to an unrelated dataset, and building baselines that rely only on the suspected shortcut attributes. Once a bias is identified, they remove it from the training data and repeat the process. In the context of PPI prediction, they identify and remove only the node degree shortcut, whereas our approach simultaneously minimizes several biases. Adding more test datasets and attribute-based baselines to our pipeline could be meaningful extensions. Interesting possible test sets could include new single studies with a non-cancer-related focus (to counter the study bias; e.g., an unbiased all-vs-all screen like \citet{eid2021systematic}), understudied organisms, or mutated or artificial sequences. The latter could also be used to study the out-of-distribution performance of the embedding models since such sequences are not included in their pretraining \citep{szymborski2026flaw}.

An important scope restriction is that this work is not a benchmarking paper: Our focus is to systematically identify biases in PPI datasets and to provide a pipeline that produces de-biased data splits with respect to the four investigated attributes for machine learning model training and evaluation. Natural follow-up question are how state-of-the-art deep learning-based PPI prediction models perform on these de-biased data splits and if they significantly outperform the Random Forest baseline utilized here. We leave such questions to future work. 

Another avenue for future work is to apply our concept of first quantifying existing shortcuts with the G-AUDIT bias detection framework and then minimizing them through optimization-based negative sampling for problems other than PPI prediction. For instance, we anticipate that it could be valuable in scenarios such as rare disease diagnosis from biomedical imaging data where there is much more data for healthy than diseased individuals and where model generalizability may be compromised by shortcuts due to data provenance or patient demographics. As long as the pool of negative candidates is sufficiently large, the approach can be extended to any binary classification problem. 

\section{Methods}\label{sec:methods}

\subsection{Datasets}\label{sec:datasets}
We selected datasets that vary along three axes relevant to shortcut formation: species scope, type of evidence, and detection modality (Table \ref{tab:ppi-datasets}). HIPPIE \citep{alanis2016hippie} and IntAct \citep{del2022intact} store interactions from dedicated experiments, the former human-only and the latter multi-species. STRING \citep{szklarczyk2025string}, in contrast, also includes non-experimentally detected interactions, and we analyze three of its channels separately to isolate their effects. PINDER \citep{kovtun2024pinder} and PDB-Dimers are derived from resolved PDB \citep{berman2000pdb} complexes rather than from interaction assays. 

HIPPIE was downloaded from \url{https://hippie-db.net/download/} (\texttt{HIPPIE-current.txt.gz}), and IntAct from \url{https://www.ebi.ac.uk/intact/download/ftp} as \texttt{intact.zip}. 
Most of IntAct's stored interactions are between human proteins (\num{618796}/\num{1082410}), followed by S. cerevisiae (\num{104961}), D. melanogaster (\num{82382}), Z.mays (\num{50969}), and A.thaliana (\num{41475}). IntAct also contains cross-species interactions, such as M. musculus–H. sapiens (\num{26586}), R. norvegicus–H. sapiens (\num{5435}), or SARS-CoV-2–H. sapiens (\num{4272}). 

The STRING database is chosen because it also contains interactions that were not experimentally measured. STRING scores each evidence channel separately and sums the scores into a combined score. A pair receives a score in the \texttt{experiments} channel if it is annotated in an experimental database like IntAct, in the \texttt{database} channel if it appears in pathway or complex databases like Reactome \citep{ragueneau2026reactome}, and in the \texttt{textmining} channel if the two proteins are co-mentioned in the literature, either through simple co-occurrence or through NLP analysis. 
STRING also distinguishes between physical and functional interactions, with functional interactions requiring only shared complex or pathway membership rather than direct contact, thereby substantially increasing the number of annotated pairs. 
Out of $\binom{\num{19311}}{2} = \num{186447705}$ possible pairwise interactions, STRING stores $\num{6737427}$, i.\,e., approximately every $27^{th}$ possible pair is annotated as interacting. 
Because each of these aspects may produce a different bias, we analyze the physical human network (\texttt{9606.protein.physical.links.full.v12.0.onlyAB.txt}) together with four of its subsets, selected by combined score and by evidence channel (Table \ref{tab:ppi-datasets}).
STRING IDs are mapped to UniProt accessions using the STRING ID mapping file (\texttt{9606.protein.aliases.v12.0.txt}), the UniProt secondary-to-primary accession mapping file (\url{https://ftp.uniprot.org/pub/databases/uniprot/current_release/knowledgebase/complete/docs/sec_ac.txt}), and the UniProt API for resolving ambiguous mappings.

We additionally investigate two structure-inferred datasets, both derived from the PDB, to examine potential differences between interaction-detection and structure-resolution databases. The first of these datasets, PINDER, was originally created to benchmark docking algorithms. It defines an interaction as any pair of chains within a biological assembly whose backbone atoms come within 10 Å of one another, and it considers all PDB complexes, including those with more than two interacting proteins. 
We downloaded the \texttt{index.parquet} file from \url{gs://pinder/2024-02/} and apply four filters. First, since we retrieve taxonomy and GO annotations from UniProt, we remove interactions with an undefined UniProt ID, which are typically small molecules (from \num{2319564} to \num{2097729} interactions). 
Second, the same pair of UniProt IDs can be annotated repeatedly within one complex, because chains with identical sequences may interact at several binding sites. \texttt{3MB8}, for instance, is a homo 6-mer forming a ring-like structure, resulting in 12 annotated \texttt{Q2HXR2-Q2HXR2} interactions in the PINDER dataset. 
Since chains sharing a UniProt ID can nonetheless differ in sequence, as in \texttt{1H5R}, where chain B of \texttt{P37744} carries two point mutations compared to chains A, C, and D, we deduplicate on PDB ID, UniProt IDs, and chain IDs rather than on UniProt IDs alone (\num{2097729} $\rightarrow$ \num{834747}). 
Third, we treat interactions as undirected and drop B–A when A–B is also present. 
Fourth, we fetch the chain sequences from the RCSB FASTA API to deduplicate on sequence (often chains have identical sequences, e.\,g., chains A, C, and D in \texttt{1H5R}). 
This leaves only \num{135398} unique interactions with sequence, taxonomy, and GO annotations, most of them between human (\num{28322}), yeast (\num{10263}), and E. coli (\num{5183}) proteins. 
37\% are self-interactions. PINDER is distributed with a predefined split that minimizes interface similarity between training, validation, and test sets. Since our aim is to characterize the dataset rather than to benchmark docking, we pool it and apply our own splitting procedure, keeping the procedure harmonized across all datasets.

We further construct a stricter structure-derived dataset, PDB-Dimers, containing biological assemblies with exactly two protein-chain instances. We do so to reduce the possibility that an observed interaction depends on additional protein chains in a higher-order complex, rather than representing an interaction that can occur between the two proteins in isolation. 
We query the PDB for structures satisfying the following criteria: \texttt{rcsb\_entry\_info.selected\_polymer\_entity\_types: Protein (only)}, \texttt{rcsb\_assembly\_info.polymer\_entity\_instance\_count: 2}, and \texttt{rcsb\_struct\_symmetry.kind: Global Symmetry}. This yields \num{97344} candidate assemblies. These criteria restrict the polymeric components of each assembly to exactly two protein chains, while allowing non-polymeric components such as ligands, cofactors, and ions. We retain such components because they do not constitute additional polymeric participants and because excluding every assembly containing a non-polymeric component would substantially reduce the available data (from \num{97344} to \num{22520} structures).
We then apply three sequence-based quality filters, excluding assemblies in which either chain is longer than \num{1250} residues (leaving \num{96766} structures), shorter than \num{50} residues (leaving \num{84504} structures), or contains more than \num{15}\% unresolved residues (leaving \num{84464} structures). Finally, sequence-identical interaction pairs are removed, and interactions B-A are dropped when A-B is present, where A and B denote protein chains (leaving \num{38758} structures).

Both structural datasets use the sequences annotated in the PDB rather than canonical UniProt sequences. Many PDB entries contain only part of the canonical sequence, either due to modifications or because the complex was made synthetically and only the peptide of interest was produced. The structural datasets, therefore, contain considerably more unique sequences than the others (Table \ref{tab:ppi-datasets}), and the same underlying protein recurs in many slightly different forms, which is why suppressing sequence similarity between splits matters for them even though its measured utility as a shortcut is low (see Results Section).

For ILP-based negative sampling, we additionally use a pool of highly replicated human non-interactions derived from the IntAct database \citep{del2022intact}. The details of the creation of this dataset are outlined in \citep{as2026negative}. The key idea is that two proteins are treated as a highly replicated negative pair if they are repeatedly not reported to interact, even though the same experiment reports each of them interacting with other proteins, demonstrating that an interaction between them would have been detected had it occurred. The original network contains $\num{99755235}$ negatives. From that, we selected negative candidates that have never been observed and have been tested at least three times.

\begin{table}[htb]
    \centering
    \caption{Datasets employed in this study.}
    \label{tab:ppi-datasets}
    \begin{tabular}{p{1.7cm}p{2cm}p{1.5cm}p{1.5cm}p{7.5cm}}
    \toprule
         Dataset Name & Version (downloaded) & Unique PPIs & Unique \newline sequences & Key Characteristics \\
         \midrule
         HIPPIE \citep{alanis2016hippie} & v3.0 (15.07.2026) & \num{1179347} & \num{29080} & Human, experimentally detected PPIs. Provides confidence scores based on the amount and quality of the experimental evidence. Aggregated from various databases. \\
         IntAct \citep{del2022intact} & Jan 2026 (10.01.2026) & \num{1082410} & \num{120021} & Multi-species, experimentally detected PPIs. Provides detailed information on the PPIs. \\
         \midrule
         STRING: physical, human \citep{szklarczyk2025string} & v12.0 (16.07.2026) & \num{644121} & \num{17348} & Human PPIs scored via different channels. Evidence for physical proximity does not only come from experiments but also from pathway and complex databases, literature text mining, and homology transfer. \\
         STRING-900 & see above & \num{37417} & \num{8032} & High-confidence PPIs subsetted from STRING: physical, human to only contain PPIs with a combined score $\geq 900$.\\
         STRING-experimental & see above & \num{414964} & \num{16624} & Subsetted from STRING: physical, human to only contain PPIs with score $>0$ in the \texttt{experiments} channel.\\
         STRING-database & see above & \num{90961} & \num{8744} & Subsetted from STRING: physical, human to only contain PPIs with score $>0$ in the \texttt{database} channel.\\
         STRING-textmining & see above & \num{105728} & \num{12852} & Subsetted from STRING: physical, human to only contain PPIs with score $>0$ in the \texttt{textmining} channel.\\
         \midrule
         PINDER \citep{kovtun2024pinder} & 2024-02 (21.07.2026) & \num{135398} & \num{91902} & Structural, multi-species dataset derived from PDB complexes. A PPI is recorded if the backbone atoms of two protein chains are closer together than $10$ Å. Customized, deduplicated version.  \\
         PDB-Dimers \citep{berman2000pdb} & 15.08.2026 & \num{38758} & \num{44706} & Structural, multi-species dataset, explicitly derived from PDB dimers instead of complexes with more than two protein units, reducing the likelihood that an observed interaction depends on additional polymers. \\
         \midrule 
        \citep{as2026negative} & 10.07.2026 & \num{25059736} & \num{13557} & Human protein pairs derived from IntAct that have been observed not to interact at least three times, despite it being possible. \\
         \bottomrule
    \end{tabular}
\end{table}

\subsection{The PPI splitting and bias analysis pipeline}\label{sec:pipeline}

To ensure a fair comparison between datasets by using the same splitting procedure and to provide the community with a tool for splitting their own PPI datasets in a bias-aware way, we implemented a dedicated PPI splitting and bias analysis pipeline. The pipeline is implemented in Nextflow (version 26.04). Its steps and the paths taken by the individual configurations are shown in Figure \ref{fig:metro-map}.
The only input required for the pipeline is a CSV file of PPIs given as UniProt accessions (required columns: \texttt{protein1}, \texttt{protein2}). 
The pipeline uses these identifiers to query the UniProt API for the amino acid sequences, the taxonomy, and the GO annotations \citep{ashburner2000go_old, aleksander2025go_new} (biological process: BP, molecular function: MF, cellular component: CC). The fetching step can also query UniParc for retired annotations and retrieve isoform-specific sequences. We exclude GO annotations whose evidence comes from the interactions themselves (evidence codes ``IPI: physical interaction'' and ``IGI: genetic interaction'').
The pipeline then offers three strategies for splitting the positives (Section \ref{sec:split-pos}) and three for sampling the negatives (Section \ref{sec:sample-neg}), which can be combined freely. It then trains and evaluates a Random Forest baseline classifier with ESM-2 or ProtT5 embeddings on the obtained data split (Section \ref{sec:train-baseline}), carries out a bias analysis via the G-AUDIT framework (Section \ref{sec:g-audit}), and generates an interactive report (Section \ref{sec:report}). The pipeline returns the resulting training, validation, and test sets with their sampled negatives, the bias metrics computed for each split, the baseline classifier's performance, and an interactive report. For the results associated with the ESM-2 embeddings, the pipeline was run in the robustness mode with three seeds for splitting (1234, 1, 2), three seeds for sampling (42, 1, 2), and three seeds for training (42, 1, 2).

\subsubsection{Splitting of the positives}\label{sec:split-pos}

\paragraph{Random split.}

The first strategy is a naive random split, in which PPIs are assigned to training, validation, and test sets at specified fractions without any further constraint. 
This approach has repeatedly been shown to produce overly optimistic results because models can memorize proteins seen during training rather than extract binding-relevant features from their sequences \citep{park2012flaws, eid2021systematic, hamp2015more, szymborski2022rapppid, bernett2024cracking}. 
We include it not as a viable option but as an upper bound on the performance a baseline can reach when every shortcut is available, against which the mitigations are measured, and it lets us quantify which biases drive that inflation. 

\paragraph{KaHIP split.}

The remaining two strategies are designed to minimize sequence similarity between splits, as suggested by \citet{hamp2015more} and \citet{bernett2024cracking}. 
To calculate sequence similarities, a custom BLAST \citep{altschul1990blast} database is created with \texttt{makeblastdb} (version 2.16.0), using only the sequences from the proteins occurring in the input CSV. 
Then, \texttt{blastp} is used for an all-against-all query (\texttt{max\_hsps} $1$) to weight a similarity graph. Because raw bit scores grow with sequence length, we define a length-normalized bit score as
\begin{equation*}
    bitscore_{norm}(p,q) = \frac{bitscore(p,q)}{min(length(p), length(q))}
\end{equation*}
which we use in all KaHIP runs. 
The graph is written in METIS format \citep{karypis1997metis}, as this is the required input format for the KaHIP \citep{sandersschulzkahip} graph partitioning tool (version 3.25). 

For the first similarity-reducing implementation, we follow the gold-standard split approach proposed by \citet{bernett2024cracking}. For this, we run KaHIP KaFFPa (Karlsruhe Fast Flow Partitioner, solves a Max-Flow Min-Cut problem) with $k=3$ to partition the proteome into three parts. Hence, sequence similarity (edge weights) is minimized among the three resulting blocks. We then sort the input PPIs into their splits, retaining a PPI $(p,q)$ only if both $p$ and $q$ are part of the same block ($\mathit{INTRA}_0$, $\mathit{INTRA}_1$, $\mathit{INTRA}_2$). The largest resulting block is defined as the training dataset, the second-largest as the validation, and the smallest as the test dataset. 

\paragraph{ILP split.}
Because the KaHIP split discards many inter-block interactions and leaves the user no control over the resulting split sizes, we implement an adaptation of DataSAIL \citep{joeres2025datasail} that formulates the assignment as an ILP. We first run KaHIP with $k=100$, so that the ILP assigns entire similarity clusters to splits rather than individual PPIs, which strongly reduces the problem size. The underlying assumption is that proteins in different clusters are dissimilar, so that assigning whole clusters keeps similar proteins on the same side of the split. However, because KaFFPa minimizes the cut between the $100$ clusters and the ILP subsequently regroups them into three splits based solely on data loss, two rather similar clusters may still end up in different splits. We compensate for this with a redundancy-removal step (see below).
As in the KaHIP split, a PPI is retained only if the clusters of both its endpoints are assigned to the same split, and discarded otherwise. The ILP chooses the cluster-to-split assignment that minimizes this data loss, subject to each split receiving approximately its target share of the retained data (notation in Table~\ref{tab:ilp-split})

\begin{table}[htb]
\centering
\caption{Notation used for ILP-based splitting.}\label{tab:ilp-split}
\begin{tabular}{@{}ll@{}}
\toprule
Symbol & Meaning \\
\midrule
\multicolumn{2}{@{}l}{Input} \\
\midrule
PPIs & The input PPIs \\
$\sC$ & Set of $n$ KaHIP clusters (default: $n=100$) \\
\midrule
\multicolumn{2}{@{}l}{Specified, fixed parameters} \\
\midrule
$S$ & Number of splits, $S = 3$ (train, val, test) \\
$f_s$ & Target fraction of split $s$, $\sum_{s=1}^{S} f_s = 1$ \\
$\varepsilon$ & Allowed fractional deviation from $f_s$, default $0.05$ \\
\midrule
\multicolumn{2}{@{}l}{Pre-computed from input} \\
\midrule
$k_{ii}$ & Number of PPIs with both endpoints in cluster $i$  \\
$k_{ij}$, $i<j$ & Number of PPIs with one endpoint in cluster $i$, the other in cluster $j$ \\
$\sP$ & $\{(i,j) : i<j,\ k_{ij} > 0\}$, the cluster pairs with at least one cross-cluster PPI \\
\midrule
\multicolumn{2}{@{}l}{Decision variables} \\
\midrule
$x_{s,c} \in \{0,1\}$ & Cluster $c$ is assigned to split $s$, $s = 1,\dots,S,\ c = 1,\dots,n$ \\
$z_{s,(i,j)} \in \{0,1\}$ & Clusters $i,j$ are both assigned to split $s$, $s = 1,\dots,S,\ (i,j) \in \sP$ \\
\bottomrule
\end{tabular}
\end{table}

The constants $k_{ii}$ and $k_{ij}$ are computed once, up front, from the PPI CSV and the KaHIP cluster assignment, counting only PPIs in which both endpoints have a sequence and a cluster assignment. A PPI whose endpoints fall in the same cluster ($k_{ii}$) is never at risk of being discarded. Note that the target fractions $f_s$ refer to shares of retained PPIs, not of proteins. 
$z_{s,(i,j)}$ is an auxiliary variable linearizing the product $x_{s,i}\cdot x_{s,j}$. It is introduced only for cluster pairs that actually have cross-cluster PPIs at stake ($\sP$). The following problem formulation is used to minimize the data loss:

\begin{align}
\min_{x,z} \quad & \sum_{(i,j) \in \sP} k_{ij} \cdot \max_{s} \big(x_{s,i} - x_{s,j}\big) \notag\\
\text{s.t.} \quad
& \sum_{s=1}^{S} x_{s,c} = 1 & \forall\, c \tag{C1}\\
& z_{s,(i,j)} \le x_{s,i},\quad z_{s,(i,j)} \le x_{s,j},\quad z_{s,(i,j)} \ge x_{s,i}+x_{s,j}-1 & \forall\, s,\ (i,j)\in \sP \tag{C2}\\
& (1-\varepsilon)\, f_s\, T(\vx,\vz) \;\le\; \sum_{i} k_{ii} x_{s,i} + \sum_{(i,j)\in \sP} k_{ij} z_{s,(i,j)} & \forall\, s \tag{C3}\\
& x_{s,c} \in \{0,1\},\quad z_{s,(i,j)} \in \{0,1\} \notag
\end{align}

For each $(i,j)\in\sP$, the corresponding summand of the objective function equals $k_{ij}$ if the clusters $i$ and $j$ are assigned to different clusters and $0$ otherwise. The constraint (C1) assigns every cluster to exactly one split. (C2) linearizes $x_{s,i}\cdot x_{s,j}$, the indicator for ``the PPIs between cluster $i$ and cluster $j$ are retained because both clusters landed in split $s$''.
(C3) requires each split to receive at least a $(1-\varepsilon)$ fraction of its target share $f_s$ of the retained PPIs, where
\begin{equation*}
T(\vx,\vz) \;=\; \underbrace{\sum_{i=1}^{n} k_{ii}}_{\text{always retained}}
\;+\; \sum_{(i,j)\in \sP} k_{ij} \sum_{s=1}^{S} z_{s,(i,j)}
\end{equation*}
is the total number of PPIs retained under the assignment $(\vx,\vz)$. 
The problem is built and solved via CVXPY (version 1.9.3). A time limit for the solver can be set. If the solver reaches the time limit before proving optimality, it returns a feasible but possibly suboptimal cluster assignment. For our experiments, the time limit was set to 2 hours, and the GUROBI solver (version 13.0.3, academic license) was used. 

\paragraph{Redundancy removal after similarity-based splits.}
Since KaFFPa partitions heuristically, and since the ILP regroups clusters without regard to inter-cluster similarity, neither similarity-based strategy fully separates the splits by sequence identity. We therefore run CD-HIT-2D~\citep{li2006cdhit} (40\% identity, word size 2) after splitting, for both strategies, and remove a PPI from the validation or test set if either of its interactors shares $\geq 40\%$ sequence identity with a training sequence.

\subsubsection{Sampling of the negatives}\label{sec:sample-neg}

\paragraph{Uniformly random sampling.}

The simplest sampler draws two proteins uniformly at random from those occurring in the positive set of the respective split. A candidate pair is retained if it does not occur among the positives. 
Negative self-interactions are allowed (though unlikely).
We use this sampler in two places. First, together with the random split, it produces the maximally biased setting, in which positives and negatives differ systematically in degree, since a protein's positive degree carries no influence on how often it is drawn as a negative. 
Second, it generates the realistic test set in every other configuration, in which we sample ten times as many negatives as positives to reflect that two arbitrary proteins are far more likely not to interact than to interact. We choose uniform sampling because true negatives are not expected to follow any specific distribution.

\paragraph{Degree-preserving sampling.}
To prevent a model from separating positives from negatives by degree alone \citep{chatterjee2023aibind}, negatives can instead be drawn such that the negative degree distribution approximately preserves the positive degree distribution \citep{Yu2010-lz,eid2021systematic,bernett2024cracking}. Here, we use the approach from \cite{bernett2024cracking} where negatives are sampled from a pool in which each protein appears as often as it does in the positive set. 
Each protein then receives, in expectation, as many negative as positive annotations.

\paragraph{ILP-based negative sampling.}
Topology is not the only shortcut that an uninformed negative sampling strategy can introduce. 
The pipeline, therefore, also offers a bias-aware negative sampler that selects negatives by solving an ILP that explicitly matches several statistical properties of the negative set to those of the positive set. 
While the ILP can be adapted to include additional or different bias terms, we focus on up to four sources of potential bias in this study: (i) whether the two sets contain a similar number of self-interactions, (ii) whether each protein has a similar degree in both sets, (iii) whether the number of interactions between each (possibly identical) pair of taxa is similar across the two sets, and (iv) whether interacting proteins have a higher mean GO biological-process Jaccard index than non-interacting proteins.

Candidates are drawn from all protein pairs not present in the positive set, optionally restricted to a user-supplied candidate network such as the highly replicated negatives. 
To speed up computation, we subsample the pool to a size four times that of the positive network. The influence of smaller or larger pools is displayed in Figures \ref{fig:pool_size_obj} and \ref{fig:pool_go_rf}.
Uniform subsampling would frequently render the problem infeasible, since pairs relevant to a given bias term are rare among all possible pairs and can be missed by chance. We therefore construct the pool by stratified sampling. 
If the self-interaction term is active, every protein without a positive self-interaction has its self-pair included unconditionally. The remaining budget is allocated proportionally across same-species and cross-species strata, using the same-species fraction of the positive set as the target ratio. 
Within each stratum, pairs with a nonzero GO-BP Jaccard index are drawn preferentially, as they would otherwise rarely appear. If a stratum's population falls short of its quota, the shortfall is filled from the unrestricted pool. Throughout, proteins are drawn in proportion to their positive degree rather than uniformly, so that low-degree proteins are not allotted candidates they cannot use while high-degree proteins receive enough to fill their larger budgets.

\begin{table}[htb]
\centering
\caption{Notation used for ILP-based negative sampling.}\label{tab:ilp-sampling}
\begin{tabularx}{\textwidth}{@{}l| X@{}}
\toprule
Symbol & Meaning \\
\midrule
\multicolumn{2}{@{}l}{Input} \\
\midrule
$\sC$ & Candidate pool of non-interacting protein pairs, $|\sC| = n$ \\
\midrule
\multicolumn{2}{@{}l}{Specified, fixed parameters} \\
\midrule
$\lambda_{\text{self}}, \lambda_{\text{deg}}, \lambda_{\text{tax}}, \lambda_{\text{jac}}$ & Weights on the individual bias terms \\
$\xi$ & Degree-cap slack, default $5$ \\
\midrule
\multicolumn{2}{@{}l}{Pre-computed from input} \\
\midrule
$n_{\text{neg}}$ & Target number of negatives $= n_{\text{pos}}$ \\
$\sC_{\text{self}} \subset \sC$ & Subset of negative self-interaction candidates \\
$s^+$ & Total number of self-interactions in the positive set \\
$d^+_p$ & Degree of protein $p$ in the positive set (a self-interaction contributes $1$, not $2$) \\
$m^+_{t,t'}$ & Number of PPIs between taxa $t$ and $t'$, $t \le t'$ in the positive set \\
$\bar{J}^+$ & Mean GO-BP Jaccard index among the positives \\
$c_p$ & Dynamic weight reflecting that a low-degree mismatch is worse than a high-degree mismatch. $c_p = 1/\log(1+d^+_p)$ \\
$\gamma_{t,t'}$ & Dynamic weight for taxon pairs, analogous to $c_p$. $\gamma_{t,t'} = \begin{cases} 1/\log(1+m^+_{t,t'}) & m^+_{t,t'} > 0 \\ 1/\log 2 & m^+_{t,t'} = 0 \end{cases}$\\
$U_{\text{self}}$ & Worst-case value of $\tau$: $\max \Bigl(s^+, \ \lvert\{(p_1, p_2) \in \sC\ :\  p_1=p_2\}\rvert - s^+\Bigr)$\\
$U_{\text{deg}}$ & \makecell[cl]{Worst-case value of $u_p$: \\ $\sum_p c_p \cdot \max \Bigl(d^+_p, \ \lvert\{(p_1, p_2) \in \sC\ :\ (p_1=p \lor p_2=p)\}\rvert - d^+_p\Bigr)$} \\
$U_{\text{tax}}$ & \makecell[cl]{Worst-case value of $v_{t, t'}$: \\ $\sum_{t, t'} \gamma_{t, t'} \cdot \max \Bigl(m^+_{t, t'}, \ \lvert\{(p_1, p_2) \in \sC\ :\ (p_1\in t \land p_2\in t') \lor (p_1\in t' \land p_2\in t)\}\rvert - m^+_{t, t'}\Bigr)$} \\$U_{\text{jac}}$ & Worst-case value of $\zeta$: $\max \Bigl(\bar{J}^+, 1-\bar{J}^+\Bigr)$\\
\midrule
\multicolumn{2}{@{}l}{Decision variables and dependent variables} \\
\midrule
$x_c \in \{0,1\}$ & Decision variable: candidate $c \in \sC$ is selected as a negative \\
$s^-(\vx)$ & Total number of negative self-interactions, $s^-(\vx) = \sum_{c \in \sC_{\text{self}}} x_c$ \\
$d^-_p(\vx)$ & Negative degree of protein $p$, $d^-_p(x) = \sum_{c \ni p} x_c$ \\
$m^-_{t,t'}(\vx)$ & Number of negative PPIs between taxa $t$ and $t'$, $m^-_{t,t'}(\vx) = \sum_{c \in \sC_{t,t'}} x_c$ \\
$J_c$ & GO-BP Jaccard index of candidate $c$'s two proteins, $J_c \in [0,1]$ \\
$\tau, u_p, v_{t,t'}, \zeta \geq 0$ & Auxiliary deviation variables for the self-loop, degree, taxon-pair, and Jaccard-mean terms \\
\bottomrule
\end{tabularx}
\end{table}

Using the notation introduced in Table~\ref{tab:ilp-sampling}, we phrase the negative sampling problem as the following ILP:
\begin{align*}
\min_{\vx,\vu,\vv,\tau,\zeta} \quad
  & \lambda_{\text{self}} \frac{1}{U_{\text{self}}}\tau
  \;+\; \lambda_{\text{deg}} \sum_{p} \frac{1}{U_{\text{deg}}}c_p\, u_p
  \;+\; \lambda_{\text{tax}} \sum_{t \le t'} \frac{1}{U_{\text{tax}}} \gamma_{t,t'}\, v_{t,t'}
  \;+\; \lambda_{\text{jac}} \frac{1}{U_{\text{jac}}}\zeta \\
\text{s.t.} \quad
& \sum_{c \in \sC} x_c = n_{\text{neg}} \tag{C1} \\
& \sum_{c \ni p} x_c \;\leq\; (1+\xi)\, d^+_p & \forall\, p \tag{C2} \\
& \tau \geq s^-(\vx) - s^+, \qquad \tau \geq  s^+ - s^-(\vx) \tag{C3} \\
& u_p \geq d^-_p(\vx) -  d^+_p, \qquad u_p \geq  d^+_p - d^-_p(\vx) & \forall\, p \tag{C4} \\
& v_{t,t'} \geq m^-_{t,t'}(\vx) -  m^+_{t,t'}, \qquad v_{t,t'} \geq  m^+_{t,t'} - m^-_{t,t'}(\vx) & \forall\, t \le t' \tag{C5} \\
& \zeta \geq \frac{1}{n_{\text{neg}}}\sum_{c} J_c\, x_c - \bar{J}^+, \qquad
  \zeta \geq \bar{J}^+ - \frac{1}{n_{\text{neg}}}\sum_{c} J_c\, x_c \tag{C6} \\
& x_c \in \{0,1\}, \qquad u_p, v_{t,t'}, \tau, \zeta \geq 0 \notag
\end{align*}

The constants $U_{\text{self}}$, $U_{\text{deg}}$, $U_{\text{tax}}$, and $U_{\text{jac}}$ rescale each bias term into $[0,1]$ by dividing through the worst-case deviation of the corresponding variable. Because the terms share candidate pairs, no two can attain their worst case simultaneously, so these constants are upper bounds on the achievable deviation. The weights $\lambda_{\text{deg}}, \lambda_{\text{tax}}, \lambda_{\text{self}}, \lambda_{\text{jac}} \ge 0$ control which terms are active and how strongly they contribute. The dynamic weights $c_p$ and $\gamma_{t,t'}$ down-weight high-degree proteins and common taxon pairs, so that a mismatch on a rarely observed protein or taxon pair is penalized more heavily than the same absolute mismatch on a frequent one. 

The constraints serve the following purposes: (C1) fixes the number of selected negatives to $n_{\text{neg}}$. (C2) caps each protein $p$'s negative degree at $(1+\xi)$ times its positive degree $d^+_p$. This hard cap is necessary because (C4) only penalizes the aggregate deviation across all proteins. On its own, it cannot prevent a small number of proteins from absorbing the entire positive/negative degree-mass mismatch. (C3) linearizes $\tau = |s^-(\vx) -  s^+|$, matching the number of self-interactions in the negative set to the number in the positive set, without constraining which specific self-pairs are chosen. Since $\tau$ is minimized in the objective and bounded below by both $s^-(\vx) - s^+$ and $s^+ - s^-(x)$, the solver drives $\tau$ to exactly this absolute deviation at the optimum. (C4) linearizes the per-protein degree deviation $u_p = |d^-_p(\vx) -  d^+_p|$. (C5) linearizes the deviation $v_{t,t'} = |m^-_{t,t'}(\vx) -  m^+_{t,t'}|$ for each unordered taxon pair $(t,t')$, matching the global distribution of interactions across species pairs independent of which specific proteins are involved. (C6) linearizes $\zeta = \big|\frac{1}{n_{\text{neg}}}\sum_e J_e x_e - \bar{J}^+\big|$, matching the mean GO-BP Jaccard index of the selected negatives to that of the positives. Unlike (C3)–(C5), which match counts within discrete groups, (C6) matches a continuous statistic averaged over the entire candidate pool.

\subsubsection{Training the baseline model}\label{sec:train-baseline}

To assess what a simple machine learning baseline can achieve on the resulting data splits, we first compute sequence embeddings for each protein. The pipeline currently supports ESM-2-650M \citep{lin2023esm2} and ProtT5 \citep{elnaggar2022prott5} embeddings, with optional GPU scheduling if specified by the user. Both models return per-residue embeddings, which we average across residues to obtain fixed-length, per-protein embeddings.

Using the concatenated embeddings of the two interactors, we train a Random Forest classifier (scikit-learn v1.9.1) with $100$ trees and \texttt{max\_samples} set to 0.2. 
We tune the maximum tree depth over $\{5, 10, 30\}$, selecting the value with the highest AUROC on the validation set.
We then report AUROC, AUPR, F1, MCC, precision, recall, and accuracy on both test sets (balanced and realistic).

\subsubsection{Bias analysis}\label{sec:g-audit}

The bias analysis is adopted from the G-AUDIT framework of \citet{drenkow2025gaudit}. The authors propose a way to quantify the extent to which a certain attribute $\va$ that is not part of the input data $\mX$ can serve as shortcut for predicting the output label $\vy$. For this, they define two measures: the detectability of $\va$ from $\mX$ and the utility of $\va$ for $\vy$, i.\,e., how informative $\va$ is about the label. Only if it is easy to infer $\va$ from $\mX$ and if $\va$ carries knowledge about $\vy$, the attribute can be considered a shortcut risk. 

Whether the attribute is actually a shortcut depends on the attribute and the application case. Returning to the introductory examples, suppose we want to classify photos of cows, dolphins, and owls. Consider the attribute ``number of green pixels'': it is easy to detect from $\mX$ (detectability) and, if most cow photos happen to be taken in meadows, it is also predictive of $\vy$ (utility). Yet this is the kind of shortcut we want to avoid, since a cow photographed on a beach would confuse the model, revealing that it never really learned ``cow'' in the first place. In contrast, the attribute ``black-and-white patches in close proximity'' would also be detectable and predictive, but since many cows genuinely do have such patches, this is a legitimate, desired signal rather than a shortcut.

In \citet{drenkow2025gaudit}, utility is measured as the mutual information $MI(\va;\vy) = H(\vy) - H(\vy|\va)$. Since the upper bound of $\mathrm{MI}$ depends on the entropies of the variables involved, raw values are not comparable across attributes, so we report the normalized mutual information
\begin{equation*}
\mathrm{NMI}(\va;\vy) = \frac{\mathrm{MI}(\va;\vy)}{\sqrt{H(\va)H(\vy)}}\,,
\end{equation*}
which is bounded in $[0,1]$.
We compute $MI(\va;\vy)$ with scikit-learn, which directly estimates mutual information via a $k$-nearest neighbors method when a continuous variable is involved, rather than computing entropies first, so $H(\vy)$ and $H(\va)$ have to be computed separately. Both can be obtained exactly for $\vy$ and discrete $\va$. 
For continuous $\va$, we compute $H(\va)$ from a 10-bin histogram. 
Since $MI(\va;\vy)$ and $H(\va)$ are thus estimated by different methods, their ratio is not guaranteed to stay within $[0,1]$, so we clip $NMI(\va;\vy)$ at $1.0$. The reported values are hence not normalized mutual information in the strict sense.

\citet{drenkow2025gaudit} measure detectability by fitting a model to predict $\va$ from $\mX$. For simplicity, we use scikit-learn's Ridge regressor ($\alpha = 1.0$, Cholesky solver, at most 1000 iterations) on the concatenated embeddings of the two interactors, and compute the Spearman correlation between $\va$ and its predictions. We use the same regressor for the binary attributes as for the continuous ones to keep the measure comparable across attributes. The regressor is deliberately fit and evaluated on the same data, since the question is whether the information about $\va$ is present in $\mX$ at all, not whether a predictor of $\va$ generalizes. Because the regressor is linear, the reported detectability is a lower bound.

We compute detectability and utility separately for the training set, the validation set, and the balanced and realistic test sets. This lets us assess whether a shortcut is present in the training data and could therefore be learned, whether it persists into the validation set and could therefore be reinforced during model selection, and whether it persists into the test sets and could therefore inflate or bias reported test performance.

We investigate the following attributes $\va$: 
\begin{itemize}
    \item \textbf{Self-interactions}: Indicator variable $\1_\mathrm{p_1 = p_2}$.
    \item \textbf{Same species}: Indicator variable $\1_\mathrm{t_1 = t_2}$, where $t_1$ and $t_2$ are the taxa annotated for $p_1$ and $p_2$. 
    \item \textbf{Functional relatedness measured by GO-BP}: The Jaccard index of the GO-BP term sets $\sS_1$ and $\sS_2$ annotated for $p_1$ and $p_2$.
    \item \textbf{Functional relatedness measured by GO-MF}: Analogous to GO-BP.
    \item \textbf{Functional relatedness measured by GO-CC}: Analogous to GO-BP. 
    \item \textbf{Sequence similarity}: The percentage of identical residues between $p_1$ and $p_2$ (BLAST output), normalized to $[0, 1]$. Self-interactions are excluded to separate the similarity signal from the identity signal.
    \item \textbf{Embedding similarity}: The cosine similarity between the protein embeddings of $p_1$ and $p_2$. Since the ridge regressor takes the concatenated embeddings as input, this attribute is by construction always highly detectable.
    \item \textbf{Topology}: The average degree ratio of the two proteins, where a protein's degree ratio is $r(p) = d^+(p)/(d^+(p)+d^-(p))$ \citep{chatterjee2023aibind}. For a pair, the attribute is the mean of the two ratios if both proteins occur in training and the ratio of the single occurring protein otherwise. Pairs in which neither protein occurs in training are excluded, though these are rare in practice, since the random split leaves almost all proteins represented in the training set. We compute this attribute only for the maximally biased setting, as all other configurations are constructed to have no train-test protein overlap, which leaves the attribute without a basis.
\end{itemize}

\subsubsection{Report}\label{sec:report}

In the end, the pipeline renders an interactive, custom MultiQC \citep{ewels2016multiqc} report by collecting individual reports from each pipeline step. This includes general statistics such as split sizes, the number of discarded PPIs, bias analysis plots, classifier performance, ILP residuals, and a sequence similarity heatmap between the splits as a sanity check.

\section{Code and data availability}

\publiconly{The code is available at \url{https://github.com/bionetslab/ppi-splitting-pipeline}, version 1.1.0. All associated data and results, including precomputed dataset splits, are available on figshare at \url{https://doi.org/10.6084/m9.figshare.33407437} (version 2).}

\anonymousonly{The code was uploaded to TMLR as a zip file. All associated data and results including precomputed dataset splits can be downloaded from the private figshare link \url{https://figshare.com/s/f0138352c081f6c397cf}. After the double-blind peer-review process, we will replace this paragraph with the link to the GitHub repository and the public figshare DOI.}

\section{Acknowledgements}

\publiconly{
The authors would like to thank Pascal Iversen for helpful discussions and feedback. This research was enabled by the Klaus Tschira Stiftung (KTS, project 00.003.2024).}

\anonymousonly{The text in this section has been removed for the double-blind peer-review process.}

\section{Broader impact statement}

All data used here come from public biological databases (UniProt, IntAct, STRING, PDB, Gene Ontology), which contain reference sequences and annotations rather than individual-level human data, so no personally identifiable information is involved.
The intended benefit is more informative benchmarks, since inflated results misdirect methodological effort and waste laboratory resources when used to prioritize experimental validation. The main risk is over-interpretation: treating a dataset as unbiased after a single mitigation, or reading a chance-level baseline as evidence that the task is unlearnable.
A potential indirect risk lies in the dual-use downstream applications of a PPI prediction model (if it now were to function well enough), where it could, e.\,g., predict host-pathogen interactions.

\bibliography{main}
\bibliographystyle{tmlr}

\clearpage
\appendix
\setcounter{figure}{0}
\renewcommand{\thefigure}{A\arabic{figure}}
\setcounter{table}{0}
\renewcommand{\thetable}{A\arabic{table}}
\section{Additional results for ESM-2 embeddings}\label{appendix-esm2}

\begin{figure}[htb]
    \centering
    \includegraphics[width=\linewidth]{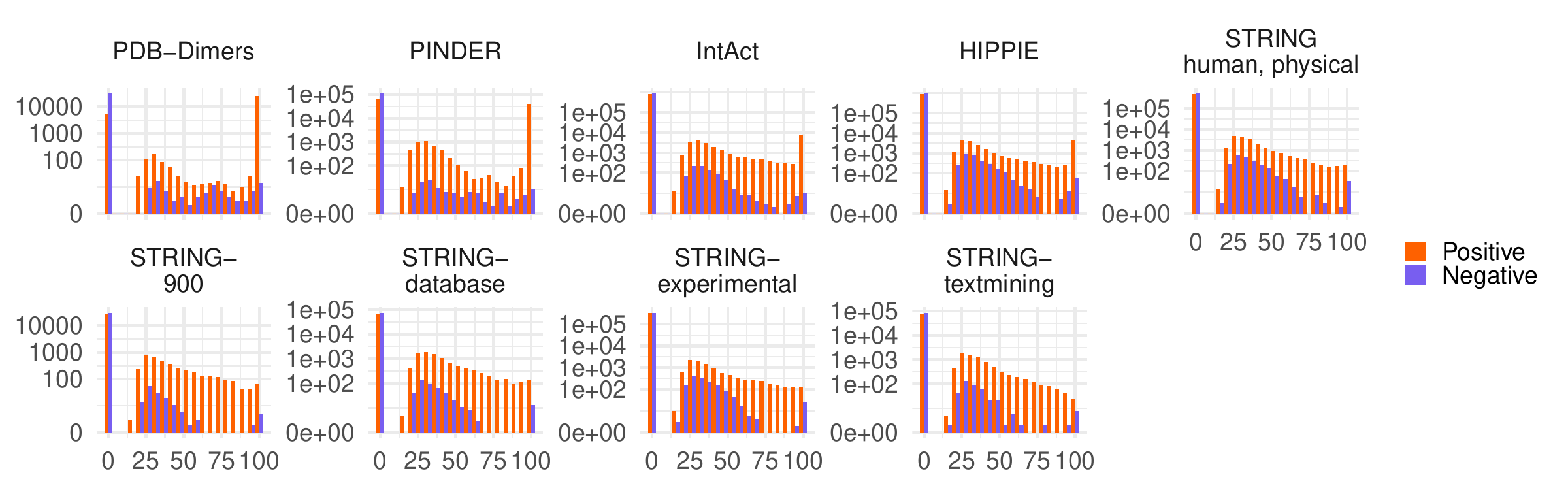}
    \caption{Percentage of identical residues (BLAST pident output) between the interaction candidates of the training datasets for the splitting/sampling combination 1234/42. The y-axis is log-scaled. }
    \label{fig:similarity}
\end{figure}

{\small\setlength{\tabcolsep}{3pt}

\begin{longtable}[t]{>{\raggedright\arraybackslash}p{2.3cm}llllllll}
\caption{Mean ± standard deviation of the Random Forest performance metrics across datasets, splitting strategies, and test sets. The results are aggregated across 27 seed combinations: 3 for splitting the dataset, 3 for sampling negatives, and 3 for training the Random Forest. \textit{(table continued on next two pages)}}\\
\toprule
Category & Test set & AUROC & AUPRC & F1 & MCC & Precision & Recall & Accuracy\\
\midrule
\endfirsthead
\caption[]{Mean ± standard deviation of the Random Forest performance metrics across datasets, splitting strategies, and test sets. The results are aggregated across 27 seed combinations: 3 for splitting the dataset, 3 for sampling negatives, and 3 for training the Random Forest. \textit{(continued)}}\\
\toprule
Category & Test set & AUROC & AUPRC & F1 & MCC & Precision & Recall & Accuracy\\
\midrule
\endhead

\endfoot
\bottomrule
\endlastfoot
\multicolumn{9}{l}{Dataset: \textbf{PDB-Dimers}}\\
\midrule
 & balanced & 0.98 $\pm$ 0.00 & 0.98 $\pm$ 0.00 & 0.94 $\pm$ 0.00 & 0.87 $\pm$ 0.01 & 0.94 $\pm$ 0.00 & 0.94 $\pm$ 0.01 & 0.94 $\pm$ 0.00\\

\multirow[t]{-1}{*}[\normalbaselineskip]{\raggedright\arraybackslash \makecell[tl]{Maximally\\biased}} & realistic & 0.98 $\pm$ 0.00 & 0.90 $\pm$ 0.01 & 0.75 $\pm$ 0.01 & 0.74 $\pm$ 0.01 & 0.63 $\pm$ 0.01 & 0.94 $\pm$ 0.01 & 0.94 $\pm$ 0.00\\

 & balanced & 0.94 $\pm$ 0.00 & 0.95 $\pm$ 0.00 & 0.86 $\pm$ 0.02 & 0.74 $\pm$ 0.02 & 0.92 $\pm$ 0.01 & 0.81 $\pm$ 0.04 & 0.87 $\pm$ 0.01\\

\multirow[t]{-1}{*}[\normalbaselineskip]{\raggedright\arraybackslash \makecell[tl]{Similarity-\\reduced}} & realistic & 0.93 $\pm$ 0.00 & 0.72 $\pm$ 0.02 & 0.58 $\pm$ 0.03 & 0.55 $\pm$ 0.02 & 0.46 $\pm$ 0.04 & 0.81 $\pm$ 0.04 & 0.89 $\pm$ 0.02\\

 & balanced & 0.99 $\pm$ 0.00 & 0.99 $\pm$ 0.00 & 0.97 $\pm$ 0.01 & 0.93 $\pm$ 0.02 & 0.95 $\pm$ 0.02 & 0.98 $\pm$ 0.01 & 0.96 $\pm$ 0.01\\

\multirow[t]{-1}{*}[\normalbaselineskip]{\raggedright\arraybackslash ILP split} & realistic & 0.99 $\pm$ 0.00 & 0.94 $\pm$ 0.03 & 0.80 $\pm$ 0.07 & 0.80 $\pm$ 0.06 & 0.68 $\pm$ 0.09 & 0.98 $\pm$ 0.01 & 0.95 $\pm$ 0.03\\

 & balanced & 0.79 $\pm$ 0.01 & 0.80 $\pm$ 0.01 & 0.73 $\pm$ 0.01 & 0.35 $\pm$ 0.04 & 0.62 $\pm$ 0.02 & 0.91 $\pm$ 0.03 & 0.66 $\pm$ 0.02\\

\multirow[t]{-1}{*}[\normalbaselineskip]{\raggedright\arraybackslash \makecell[tl]{ILP split +\\ILP sampling}} & realistic & 0.80 $\pm$ 0.01 & 0.56 $\pm$ 0.03 & 0.21 $\pm$ 0.01 & 0.13 $\pm$ 0.01 & 0.12 $\pm$ 0.00 & 0.91 $\pm$ 0.03 & 0.36 $\pm$ 0.03\\

\midrule
\multicolumn{9}{l}{Dataset: \textbf{PINDER}}\\
\midrule
 & balanced & 0.95 $\pm$ 0.00 & 0.95 $\pm$ 0.00 & 0.88 $\pm$ 0.00 & 0.76 $\pm$ 0.00 & 0.87 $\pm$ 0.00 & 0.89 $\pm$ 0.00 & 0.88 $\pm$ 0.00\\

\multirow[t]{-1}{*}[\normalbaselineskip]{\raggedright\arraybackslash \makecell[tl]{Maximally\\biased}} & realistic & 0.95 $\pm$ 0.00 & 0.75 $\pm$ 0.00 & 0.56 $\pm$ 0.00 & 0.55 $\pm$ 0.00 & 0.41 $\pm$ 0.00 & 0.89 $\pm$ 0.00 & 0.87 $\pm$ 0.00\\

 & balanced & 0.93 $\pm$ 0.03 & 0.94 $\pm$ 0.03 & 0.87 $\pm$ 0.03 & 0.75 $\pm$ 0.05 & 0.88 $\pm$ 0.02 & 0.86 $\pm$ 0.06 & 0.87 $\pm$ 0.03\\

\multirow[t]{-1}{*}[\normalbaselineskip]{\raggedright\arraybackslash \makecell[tl]{Similarity-\\reduced}} & realistic & 0.93 $\pm$ 0.03 & 0.72 $\pm$ 0.14 & 0.57 $\pm$ 0.05 & 0.56 $\pm$ 0.05 & 0.43 $\pm$ 0.06 & 0.86 $\pm$ 0.06 & 0.88 $\pm$ 0.02\\

 & balanced & 0.95 $\pm$ 0.02 & 0.95 $\pm$ 0.02 & 0.87 $\pm$ 0.03 & 0.76 $\pm$ 0.05 & 0.93 $\pm$ 0.01 & 0.81 $\pm$ 0.06 & 0.88 $\pm$ 0.03\\

\multirow[t]{-1}{*}[\normalbaselineskip]{\raggedright\arraybackslash ILP split} & realistic & 0.95 $\pm$ 0.02 & 0.79 $\pm$ 0.06 & 0.68 $\pm$ 0.03 & 0.66 $\pm$ 0.03 & 0.60 $\pm$ 0.03 & 0.81 $\pm$ 0.06 & 0.93 $\pm$ 0.01\\

 & balanced & 0.74 $\pm$ 0.04 & 0.76 $\pm$ 0.02 & 0.69 $\pm$ 0.04 & 0.29 $\pm$ 0.12 & 0.60 $\pm$ 0.06 & 0.82 $\pm$ 0.06 & 0.63 $\pm$ 0.07\\

\multirow[t]{-1}{*}[\normalbaselineskip]{\raggedright\arraybackslash \makecell[tl]{ILP split +\\ILP sampling}} & realistic & 0.63 $\pm$ 0.03 & 0.27 $\pm$ 0.03 & 0.17 $\pm$ 0.01 & 0.03 $\pm$ 0.04 & 0.10 $\pm$ 0.01 & 0.82 $\pm$ 0.06 & 0.29 $\pm$ 0.10\\

\midrule
\multicolumn{9}{l}{Dataset: \textbf{IntAct}}\\
\midrule
 & balanced & 0.88 $\pm$ 0.00 & 0.89 $\pm$ 0.00 & 0.82 $\pm$ 0.00 & 0.58 $\pm$ 0.00 & 0.75 $\pm$ 0.00 & 0.92 $\pm$ 0.00 & 0.78 $\pm$ 0.00\\

\multirow[t]{-1}{*}[\normalbaselineskip]{\raggedright\arraybackslash \makecell[tl]{Maximally\\biased}} & realistic & 0.88 $\pm$ 0.00 & 0.52 $\pm$ 0.00 & 0.37 $\pm$ 0.00 & 0.34 $\pm$ 0.00 & 0.23 $\pm$ 0.00 & 0.92 $\pm$ 0.00 & 0.65 $\pm$ 0.00\\

 & balanced & 0.72 $\pm$ 0.00 & 0.70 $\pm$ 0.00 & 0.65 $\pm$ 0.02 & 0.30 $\pm$ 0.01 & 0.65 $\pm$ 0.00 & 0.65 $\pm$ 0.03 & 0.65 $\pm$ 0.01\\

\multirow[t]{-1}{*}[\normalbaselineskip]{\raggedright\arraybackslash \makecell[tl]{Similarity-\\reduced}} & realistic & 0.72 $\pm$ 0.01 & 0.23 $\pm$ 0.00 & 0.28 $\pm$ 0.00 & 0.19 $\pm$ 0.00 & 0.18 $\pm$ 0.00 & 0.65 $\pm$ 0.03 & 0.66 $\pm$ 0.02\\

 & balanced & 0.69 $\pm$ 0.04 & 0.67 $\pm$ 0.04 & 0.64 $\pm$ 0.02 & 0.27 $\pm$ 0.07 & 0.63 $\pm$ 0.04 & 0.65 $\pm$ 0.04 & 0.63 $\pm$ 0.03\\

\multirow[t]{-1}{*}[\normalbaselineskip]{\raggedright\arraybackslash ILP split} & realistic & 0.74 $\pm$ 0.01 & 0.25 $\pm$ 0.02 & 0.31 $\pm$ 0.01 & 0.22 $\pm$ 0.01 & 0.20 $\pm$ 0.01 & 0.65 $\pm$ 0.04 & 0.69 $\pm$ 0.03\\

 & balanced & 0.55 $\pm$ 0.01 & 0.56 $\pm$ 0.01 & 0.55 $\pm$ 0.06 & 0.06 $\pm$ 0.02 & 0.53 $\pm$ 0.02 & 0.59 $\pm$ 0.15 & 0.53 $\pm$ 0.01\\

\multirow[t]{-1}{*}[\normalbaselineskip]{\raggedright\arraybackslash \makecell[tl]{ILP split +\\ILP sampling}} & realistic & 0.41 $\pm$ 0.04 & 0.09 $\pm$ 0.01 & 0.15 $\pm$ 0.02 & -0.09 $\pm$ 0.04 & 0.09 $\pm$ 0.01 & 0.59 $\pm$ 0.15 & 0.31 $\pm$ 0.07\\

\midrule
\pagebreak
\multicolumn{9}{l}{Dataset: \textbf{HIPPIE}}\\
\midrule
 & balanced & 0.87 $\pm$ 0.00 & 0.87 $\pm$ 0.00 & 0.80 $\pm$ 0.00 & 0.57 $\pm$ 0.00 & 0.76 $\pm$ 0.00 & 0.84 $\pm$ 0.00 & 0.78 $\pm$ 0.00\\

\multirow[t]{-1}{*}[\normalbaselineskip]{\raggedright\arraybackslash \makecell[tl]{Maximally\\biased}} & realistic & 0.86 $\pm$ 0.00 & 0.48 $\pm$ 0.00 & 0.37 $\pm$ 0.00 & 0.35 $\pm$ 0.00 & 0.24 $\pm$ 0.00 & 0.84 $\pm$ 0.00 & 0.73 $\pm$ 0.00\\

 & balanced & 0.59 $\pm$ 0.00 & 0.58 $\pm$ 0.00 & 0.47 $\pm$ 0.02 & 0.12 $\pm$ 0.00 & 0.58 $\pm$ 0.01 & 0.39 $\pm$ 0.03 & 0.56 $\pm$ 0.00\\

\multirow[t]{-1}{*}[\normalbaselineskip]{\raggedright\arraybackslash \makecell[tl]{Similarity-\\reduced}} & realistic & 0.63 $\pm$ 0.01 & 0.15 $\pm$ 0.01 & 0.22 $\pm$ 0.01 & 0.11 $\pm$ 0.01 & 0.15 $\pm$ 0.01 & 0.39 $\pm$ 0.03 & 0.73 $\pm$ 0.01\\

 & balanced & 0.58 $\pm$ 0.00 & 0.57 $\pm$ 0.01 & 0.47 $\pm$ 0.03 & 0.10 $\pm$ 0.01 & 0.57 $\pm$ 0.01 & 0.40 $\pm$ 0.05 & 0.55 $\pm$ 0.00\\

\multirow[t]{-1}{*}[\normalbaselineskip]{\raggedright\arraybackslash ILP split} & realistic & 0.61 $\pm$ 0.03 & 0.13 $\pm$ 0.02 & 0.20 $\pm$ 0.03 & 0.09 $\pm$ 0.04 & 0.14 $\pm$ 0.02 & 0.40 $\pm$ 0.05 & 0.70 $\pm$ 0.01\\

 & balanced & 0.55 $\pm$ 0.01 & 0.55 $\pm$ 0.01 & 0.54 $\pm$ 0.02 & 0.07 $\pm$ 0.01 & 0.53 $\pm$ 0.01 & 0.55 $\pm$ 0.04 & 0.53 $\pm$ 0.00\\

\multirow[t]{-1}{*}[\normalbaselineskip]{\raggedright\arraybackslash \makecell[tl]{ILP split +\\ILP sampling}} & realistic & 0.53 $\pm$ 0.03 & 0.10 $\pm$ 0.01 & 0.18 $\pm$ 0.01 & 0.03 $\pm$ 0.02 & 0.10 $\pm$ 0.01 & 0.55 $\pm$ 0.04 & 0.50 $\pm$ 0.06\\

 & balanced & 0.57 $\pm$ 0.01 & 0.57 $\pm$ 0.02 & 0.64 $\pm$ 0.01 & 0.08 $\pm$ 0.02 & 0.52 $\pm$ 0.01 & 0.85 $\pm$ 0.04 & 0.53 $\pm$ 0.01\\

\multirow[t]{-1}{*}[\normalbaselineskip]{\raggedright\arraybackslash \makecell[tl]{ILP split +\\sampling (HRNI)}} & realistic & 0.38 $\pm$ 0.02 & 0.07 $\pm$ 0.00 & 0.16 $\pm$ 0.01 & -0.10 $\pm$ 0.02 & 0.09 $\pm$ 0.00 & 0.85 $\pm$ 0.04 & 0.14 $\pm$ 0.02\\

\midrule
\multicolumn{9}{l}{Dataset: \textbf{STRING: human, physical}}\\
\midrule
 & balanced & 0.85 $\pm$ 0.00 & 0.85 $\pm$ 0.00 & 0.77 $\pm$ 0.00 & 0.54 $\pm$ 0.00 & 0.77 $\pm$ 0.00 & 0.77 $\pm$ 0.00 & 0.77 $\pm$ 0.00\\

\multirow[t]{-1}{*}[\normalbaselineskip]{\raggedright\arraybackslash \makecell[tl]{Maximally\\biased}} & realistic & 0.85 $\pm$ 0.00 & 0.47 $\pm$ 0.00 & 0.37 $\pm$ 0.00 & 0.34 $\pm$ 0.00 & 0.24 $\pm$ 0.00 & 0.77 $\pm$ 0.00 & 0.76 $\pm$ 0.00\\

 & balanced & 0.62 $\pm$ 0.02 & 0.61 $\pm$ 0.01 & 0.53 $\pm$ 0.01 & 0.16 $\pm$ 0.02 & 0.60 $\pm$ 0.02 & 0.47 $\pm$ 0.03 & 0.58 $\pm$ 0.01\\

\multirow[t]{-1}{*}[\normalbaselineskip]{\raggedright\arraybackslash \makecell[tl]{Similarity-\\reduced}} & realistic & 0.64 $\pm$ 0.02 & 0.16 $\pm$ 0.01 & 0.23 $\pm$ 0.01 & 0.12 $\pm$ 0.01 & 0.15 $\pm$ 0.01 & 0.47 $\pm$ 0.03 & 0.70 $\pm$ 0.02\\

 & balanced & 0.62 $\pm$ 0.01 & 0.62 $\pm$ 0.01 & 0.51 $\pm$ 0.03 & 0.16 $\pm$ 0.01 & 0.61 $\pm$ 0.01 & 0.45 $\pm$ 0.05 & 0.58 $\pm$ 0.00\\

\multirow[t]{-1}{*}[\normalbaselineskip]{\raggedright\arraybackslash ILP split} & realistic & 0.65 $\pm$ 0.03 & 0.16 $\pm$ 0.02 & 0.22 $\pm$ 0.02 & 0.12 $\pm$ 0.02 & 0.15 $\pm$ 0.01 & 0.45 $\pm$ 0.05 & 0.72 $\pm$ 0.02\\

 & balanced & 0.56 $\pm$ 0.01 & 0.56 $\pm$ 0.01 & 0.55 $\pm$ 0.02 & 0.08 $\pm$ 0.01 & 0.54 $\pm$ 0.01 & 0.55 $\pm$ 0.04 & 0.54 $\pm$ 0.01\\

\multirow[t]{-1}{*}[\normalbaselineskip]{\raggedright\arraybackslash \makecell[tl]{ILP split +\\ILP sampling}} & realistic & 0.55 $\pm$ 0.02 & 0.11 $\pm$ 0.01 & 0.18 $\pm$ 0.01 & 0.04 $\pm$ 0.01 & 0.10 $\pm$ 0.01 & 0.55 $\pm$ 0.04 & 0.53 $\pm$ 0.05\\

 & balanced & 0.56 $\pm$ 0.06 & 0.56 $\pm$ 0.06 & 0.65 $\pm$ 0.02 & 0.06 $\pm$ 0.06 & 0.51 $\pm$ 0.01 & 0.87 $\pm$ 0.04 & 0.52 $\pm$ 0.02\\

\multirow[t]{-1}{*}[\normalbaselineskip]{\raggedright\arraybackslash \makecell[tl]{ILP split +\\sampling (HRNI)}} & realistic & 0.37 $\pm$ 0.06 & 0.07 $\pm$ 0.01 & 0.15 $\pm$ 0.00 & -0.10 $\pm$ 0.02 & 0.08 $\pm$ 0.00 & 0.87 $\pm$ 0.04 & 0.12 $\pm$ 0.01\\

\midrule
\multicolumn{9}{l}{Dataset: \textbf{STRING-900}}\\
\midrule
 & balanced & 0.82 $\pm$ 0.00 & 0.83 $\pm$ 0.00 & 0.74 $\pm$ 0.00 & 0.46 $\pm$ 0.01 & 0.71 $\pm$ 0.01 & 0.77 $\pm$ 0.01 & 0.73 $\pm$ 0.01\\

\multirow[t]{-1}{*}[\normalbaselineskip]{\raggedright\arraybackslash \makecell[tl]{Maximally\\biased}} & realistic & 0.82 $\pm$ 0.00 & 0.45 $\pm$ 0.01 & 0.31 $\pm$ 0.00 & 0.27 $\pm$ 0.00 & 0.20 $\pm$ 0.00 & 0.77 $\pm$ 0.01 & 0.69 $\pm$ 0.00\\

 & balanced & 0.68 $\pm$ 0.01 & 0.67 $\pm$ 0.01 & 0.61 $\pm$ 0.02 & 0.26 $\pm$ 0.02 & 0.64 $\pm$ 0.01 & 0.57 $\pm$ 0.04 & 0.63 $\pm$ 0.01\\

\multirow[t]{-1}{*}[\normalbaselineskip]{\raggedright\arraybackslash \makecell[tl]{Similarity-\\reduced}} & realistic & 0.67 $\pm$ 0.01 & 0.19 $\pm$ 0.02 & 0.24 $\pm$ 0.01 & 0.15 $\pm$ 0.01 & 0.15 $\pm$ 0.01 & 0.57 $\pm$ 0.04 & 0.66 $\pm$ 0.04\\

 & balanced & 0.68 $\pm$ 0.03 & 0.65 $\pm$ 0.03 & 0.63 $\pm$ 0.02 & 0.26 $\pm$ 0.05 & 0.63 $\pm$ 0.04 & 0.65 $\pm$ 0.05 & 0.63 $\pm$ 0.02\\

\multirow[t]{-1}{*}[\normalbaselineskip]{\raggedright\arraybackslash ILP split} & realistic & 0.67 $\pm$ 0.03 & 0.17 $\pm$ 0.02 & 0.23 $\pm$ 0.02 & 0.15 $\pm$ 0.03 & 0.14 $\pm$ 0.02 & 0.65 $\pm$ 0.05 & 0.61 $\pm$ 0.06\\

 & balanced & 0.60 $\pm$ 0.03 & 0.57 $\pm$ 0.02 & 0.60 $\pm$ 0.02 & 0.14 $\pm$ 0.04 & 0.56 $\pm$ 0.02 & 0.64 $\pm$ 0.06 & 0.57 $\pm$ 0.02\\

\multirow[t]{-1}{*}[\normalbaselineskip]{\raggedright\arraybackslash \makecell[tl]{ILP split +\\ILP sampling}} & realistic & 0.65 $\pm$ 0.01 & 0.15 $\pm$ 0.01 & 0.22 $\pm$ 0.01 & 0.12 $\pm$ 0.01 & 0.13 $\pm$ 0.01 & 0.64 $\pm$ 0.06 & 0.58 $\pm$ 0.05\\

 & balanced & 0.65 $\pm$ 0.03 & 0.65 $\pm$ 0.02 & 0.65 $\pm$ 0.03 & 0.16 $\pm$ 0.03 & 0.54 $\pm$ 0.01 & 0.81 $\pm$ 0.10 & 0.56 $\pm$ 0.01\\

\multirow[t]{-1}{*}[\normalbaselineskip]{\raggedright\arraybackslash \makecell[tl]{ILP split +\\sampling (HRNI)}} & realistic & 0.56 $\pm$ 0.03 & 0.12 $\pm$ 0.01 & 0.17 $\pm$ 0.01 & 0.03 $\pm$ 0.03 & 0.09 $\pm$ 0.00 & 0.81 $\pm$ 0.10 & 0.27 $\pm$ 0.06\\

\midrule
\multicolumn{9}{l}{Dataset: \textbf{STRING-experimental}}\\
\midrule
 & balanced & 0.83 $\pm$ 0.00 & 0.84 $\pm$ 0.00 & 0.75 $\pm$ 0.00 & 0.50 $\pm$ 0.00 & 0.74 $\pm$ 0.00 & 0.76 $\pm$ 0.00 & 0.75 $\pm$ 0.00\\

\multirow[t]{-1}{*}[\normalbaselineskip]{\raggedright\arraybackslash \makecell[tl]{Maximally\\biased}} & realistic & 0.83 $\pm$ 0.00 & 0.45 $\pm$ 0.00 & 0.35 $\pm$ 0.00 & 0.31 $\pm$ 0.00 & 0.23 $\pm$ 0.00 & 0.76 $\pm$ 0.00 & 0.74 $\pm$ 0.00\\

 & balanced & 0.57 $\pm$ 0.00 & 0.57 $\pm$ 0.00 & 0.51 $\pm$ 0.01 & 0.10 $\pm$ 0.01 & 0.55 $\pm$ 0.00 & 0.48 $\pm$ 0.02 & 0.55 $\pm$ 0.00\\

\multirow[t]{-1}{*}[\normalbaselineskip]{\raggedright\arraybackslash \makecell[tl]{Similarity-\\reduced}} & realistic & 0.60 $\pm$ 0.01 & 0.13 $\pm$ 0.00 & 0.20 $\pm$ 0.00 & 0.08 $\pm$ 0.00 & 0.12 $\pm$ 0.00 & 0.48 $\pm$ 0.02 & 0.64 $\pm$ 0.02\\

 & balanced & 0.57 $\pm$ 0.01 & 0.56 $\pm$ 0.01 & 0.52 $\pm$ 0.03 & 0.09 $\pm$ 0.01 & 0.55 $\pm$ 0.01 & 0.49 $\pm$ 0.07 & 0.54 $\pm$ 0.01\\

\multirow[t]{-1}{*}[\normalbaselineskip]{\raggedright\arraybackslash ILP split} & realistic & 0.63 $\pm$ 0.03 & 0.14 $\pm$ 0.02 & 0.21 $\pm$ 0.02 & 0.10 $\pm$ 0.03 & 0.13 $\pm$ 0.01 & 0.49 $\pm$ 0.07 & 0.66 $\pm$ 0.02\\

 & balanced & 0.53 $\pm$ 0.01 & 0.53 $\pm$ 0.01 & 0.52 $\pm$ 0.01 & 0.04 $\pm$ 0.01 & 0.52 $\pm$ 0.01 & 0.52 $\pm$ 0.03 & 0.52 $\pm$ 0.01\\

\multirow[t]{-1}{*}[\normalbaselineskip]{\raggedright\arraybackslash \makecell[tl]{ILP split +\\ILP sampling}} & realistic & 0.52 $\pm$ 0.02 & 0.10 $\pm$ 0.01 & 0.17 $\pm$ 0.01 & 0.02 $\pm$ 0.02 & 0.10 $\pm$ 0.01 & 0.52 $\pm$ 0.03 & 0.52 $\pm$ 0.04\\

 & balanced & 0.58 $\pm$ 0.01 & 0.59 $\pm$ 0.01 & 0.63 $\pm$ 0.01 & 0.08 $\pm$ 0.01 & 0.52 $\pm$ 0.01 & 0.79 $\pm$ 0.06 & 0.53 $\pm$ 0.01\\

\multirow[t]{-1}{*}[\normalbaselineskip]{\raggedright\arraybackslash \makecell[tl]{ILP split +\\sampling (HRNI)}} & realistic & 0.36 $\pm$ 0.03 & 0.07 $\pm$ 0.00 & 0.15 $\pm$ 0.01 & -0.10 $\pm$ 0.03 & 0.08 $\pm$ 0.00 & 0.79 $\pm$ 0.06 & 0.17 $\pm$ 0.03\\

\midrule
\multicolumn{9}{l}{Dataset: \textbf{STRING-database}}\\
\midrule
 & balanced & 0.86 $\pm$ 0.00 & 0.86 $\pm$ 0.00 & 0.78 $\pm$ 0.00 & 0.54 $\pm$ 0.00 & 0.73 $\pm$ 0.00 & 0.84 $\pm$ 0.00 & 0.77 $\pm$ 0.00\\

\multirow[t]{-1}{*}[\normalbaselineskip]{\raggedright\arraybackslash \makecell[tl]{Maximally\\biased}} & realistic & 0.86 $\pm$ 0.00 & 0.48 $\pm$ 0.01 & 0.34 $\pm$ 0.00 & 0.32 $\pm$ 0.00 & 0.22 $\pm$ 0.00 & 0.84 $\pm$ 0.00 & 0.71 $\pm$ 0.00\\

 & balanced & 0.64 $\pm$ 0.01 & 0.63 $\pm$ 0.01 & 0.53 $\pm$ 0.02 & 0.19 $\pm$ 0.02 & 0.63 $\pm$ 0.01 & 0.46 $\pm$ 0.03 & 0.59 $\pm$ 0.01\\

\multirow[t]{-1}{*}[\normalbaselineskip]{\raggedright\arraybackslash \makecell[tl]{Similarity-\\reduced}} & realistic & 0.65 $\pm$ 0.01 & 0.16 $\pm$ 0.01 & 0.23 $\pm$ 0.01 & 0.13 $\pm$ 0.01 & 0.15 $\pm$ 0.01 & 0.46 $\pm$ 0.03 & 0.71 $\pm$ 0.02\\

 & balanced & 0.68 $\pm$ 0.02 & 0.67 $\pm$ 0.02 & 0.56 $\pm$ 0.03 & 0.25 $\pm$ 0.02 & 0.66 $\pm$ 0.02 & 0.49 $\pm$ 0.04 & 0.62 $\pm$ 0.01\\

\multirow[t]{-1}{*}[\normalbaselineskip]{\raggedright\arraybackslash ILP split} & realistic & 0.68 $\pm$ 0.01 & 0.18 $\pm$ 0.02 & 0.25 $\pm$ 0.01 & 0.15 $\pm$ 0.01 & 0.16 $\pm$ 0.01 & 0.49 $\pm$ 0.04 & 0.73 $\pm$ 0.03\\

 & balanced & 0.61 $\pm$ 0.01 & 0.59 $\pm$ 0.02 & 0.53 $\pm$ 0.03 & 0.15 $\pm$ 0.02 & 0.59 $\pm$ 0.01 & 0.49 $\pm$ 0.05 & 0.57 $\pm$ 0.01\\

\multirow[t]{-1}{*}[\normalbaselineskip]{\raggedright\arraybackslash \makecell[tl]{ILP split +\\ILP sampling}} & realistic & 0.64 $\pm$ 0.01 & 0.15 $\pm$ 0.01 & 0.22 $\pm$ 0.01 & 0.12 $\pm$ 0.01 & 0.14 $\pm$ 0.01 & 0.49 $\pm$ 0.05 & 0.68 $\pm$ 0.04\\

 & balanced & 0.66 $\pm$ 0.01 & 0.68 $\pm$ 0.02 & 0.62 $\pm$ 0.01 & 0.22 $\pm$ 0.03 & 0.60 $\pm$ 0.03 & 0.65 $\pm$ 0.06 & 0.61 $\pm$ 0.02\\

\multirow[t]{-1}{*}[\normalbaselineskip]{\raggedright\arraybackslash \makecell[tl]{ILP split +\\sampling (HRNI)}} & realistic & 0.50 $\pm$ 0.03 & 0.11 $\pm$ 0.01 & 0.16 $\pm$ 0.01 & -0.02 $\pm$ 0.03 & 0.09 $\pm$ 0.00 & 0.65 $\pm$ 0.06 & 0.36 $\pm$ 0.03\\

\midrule
\multicolumn{9}{l}{Dataset: \textbf{STRING-textmining}}\\
\midrule
 & balanced & 0.80 $\pm$ 0.00 & 0.81 $\pm$ 0.00 & 0.73 $\pm$ 0.00 & 0.45 $\pm$ 0.00 & 0.71 $\pm$ 0.00 & 0.75 $\pm$ 0.00 & 0.72 $\pm$ 0.00\\

\multirow[t]{-1}{*}[\normalbaselineskip]{\raggedright\arraybackslash \makecell[tl]{Maximally\\biased}} & realistic & 0.80 $\pm$ 0.00 & 0.40 $\pm$ 0.01 & 0.32 $\pm$ 0.00 & 0.27 $\pm$ 0.00 & 0.20 $\pm$ 0.00 & 0.75 $\pm$ 0.00 & 0.70 $\pm$ 0.00\\

 & balanced & 0.62 $\pm$ 0.03 & 0.61 $\pm$ 0.03 & 0.57 $\pm$ 0.02 & 0.16 $\pm$ 0.04 & 0.59 $\pm$ 0.03 & 0.55 $\pm$ 0.03 & 0.58 $\pm$ 0.02\\

\multirow[t]{-1}{*}[\normalbaselineskip]{\raggedright\arraybackslash \makecell[tl]{Similarity-\\reduced}} & realistic & 0.64 $\pm$ 0.02 & 0.16 $\pm$ 0.02 & 0.22 $\pm$ 0.01 & 0.12 $\pm$ 0.02 & 0.13 $\pm$ 0.01 & 0.55 $\pm$ 0.03 & 0.64 $\pm$ 0.02\\

 & balanced & 0.62 $\pm$ 0.07 & 0.61 $\pm$ 0.06 & 0.59 $\pm$ 0.06 & 0.17 $\pm$ 0.10 & 0.58 $\pm$ 0.05 & 0.61 $\pm$ 0.09 & 0.58 $\pm$ 0.05\\

\multirow[t]{-1}{*}[\normalbaselineskip]{\raggedright\arraybackslash ILP split} & realistic & 0.65 $\pm$ 0.06 & 0.17 $\pm$ 0.05 & 0.22 $\pm$ 0.04 & 0.12 $\pm$ 0.06 & 0.13 $\pm$ 0.02 & 0.61 $\pm$ 0.09 & 0.60 $\pm$ 0.07\\

 & balanced & 0.56 $\pm$ 0.02 & 0.54 $\pm$ 0.01 & 0.56 $\pm$ 0.03 & 0.08 $\pm$ 0.03 & 0.54 $\pm$ 0.01 & 0.60 $\pm$ 0.07 & 0.54 $\pm$ 0.02\\

\multirow[t]{-1}{*}[\normalbaselineskip]{\raggedright\arraybackslash \makecell[tl]{ILP split +\\ILP sampling}} & realistic & 0.61 $\pm$ 0.05 & 0.14 $\pm$ 0.03 & 0.20 $\pm$ 0.02 & 0.09 $\pm$ 0.04 & 0.12 $\pm$ 0.02 & 0.60 $\pm$ 0.07 & 0.56 $\pm$ 0.05\\

 & balanced & 0.58 $\pm$ 0.03 & 0.58 $\pm$ 0.03 & 0.65 $\pm$ 0.02 & 0.08 $\pm$ 0.02 & 0.52 $\pm$ 0.00 & 0.86 $\pm$ 0.07 & 0.53 $\pm$ 0.01\\

\multirow[t]{-1}{*}[\normalbaselineskip]{\raggedright\arraybackslash \makecell[tl]{ILP split +\\sampling (HRNI)}} & realistic & 0.49 $\pm$ 0.06 & 0.10 $\pm$ 0.02 & 0.16 $\pm$ 0.01 & -0.00 $\pm$ 0.03 & 0.09 $\pm$ 0.00 & 0.86 $\pm$ 0.07 & 0.20 $\pm$ 0.03\\*
\label{tab:all_baseline_results}
\end{longtable}

\begin{figure}[htb]
    \centering
    \includegraphics[width=0.95\linewidth]{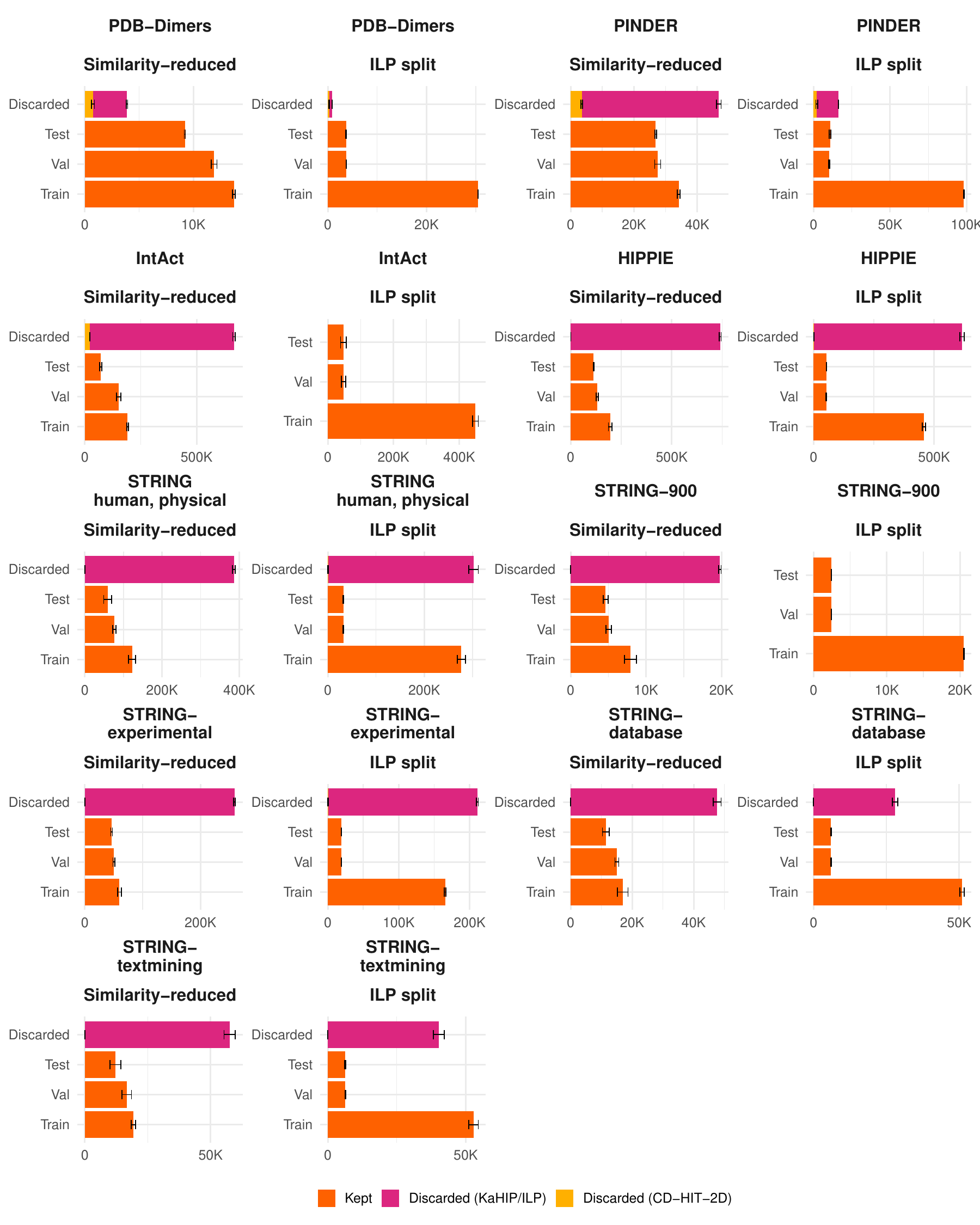}
    \caption{Mean ± standard deviation of the data retained for training, validation, and test by the two suggested splitting approaches vs. data discarded by them across the 3 splitting seeds. CD-HIT-2D mostly removes leftover similar interactions in IntAct ($\num{21656} \pm 77$ in the similarity-reduced setting), HIPPIE ($\num{2078} \pm 423$ in the similarity-reduced setting), and PINDER ($\num{3516} \pm 314$ in the similarity-reduced setting and $\num{2119} \pm 586$ in the ILP split). For the other datasets, fewer than 1000 interactions are removed.}
    \label{fig:cdhit}
\end{figure}

\begin{figure}[htb]
    \centering
    \includegraphics[width=\linewidth]{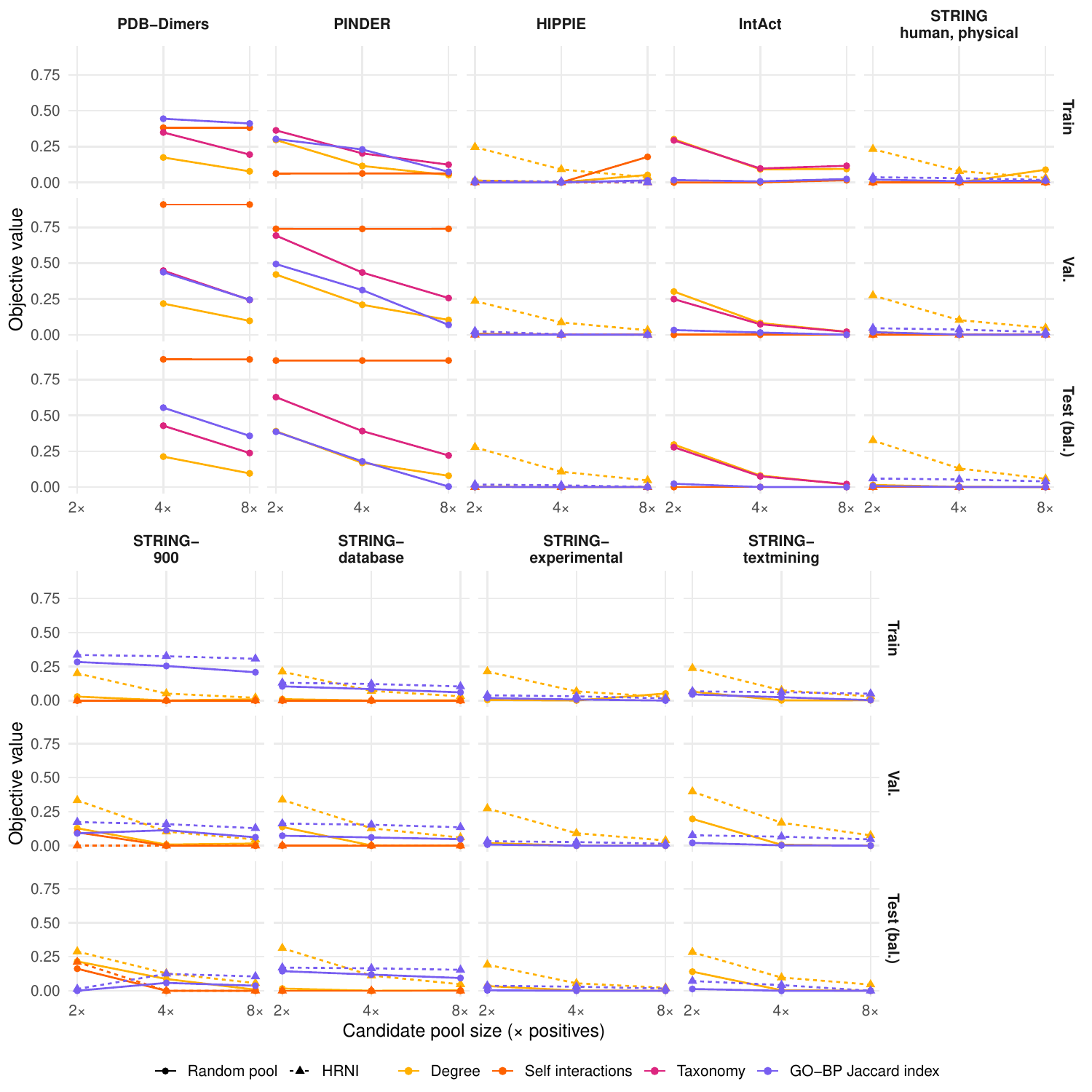}
    \caption{Influence of the negative pool size on the objective value. Because the sampling ILP otherwise becomes too large, we do not solve the ILP from the whole set of possible negatives (the complement or the HRNI) but subsample this set to a candidate pool ($4\timesend$ the number of positives in the main text). In this figure, the effect of a smaller or larger candidate pool size on the objective value is shown (split/sampling seed combination 1234/42). The ILP solution was always optimal for the given pool. For the small $2\timesend$ pool, no feasible solution was found for  PDB-Dimers. In general, a larger pool leads to smaller residual objective values. Exceptions are bias categories that cannot be minimized even from the full pool, i.\,e., the self-interactions for the structural datasets and the GO-BP overlap for STRING-900 and STRING-database, especially when sampling from the HRNIs.}
    \label{fig:pool_size_obj}
\end{figure}

\begin{figure}[htb]
    \centering
    \includegraphics[width=\linewidth]{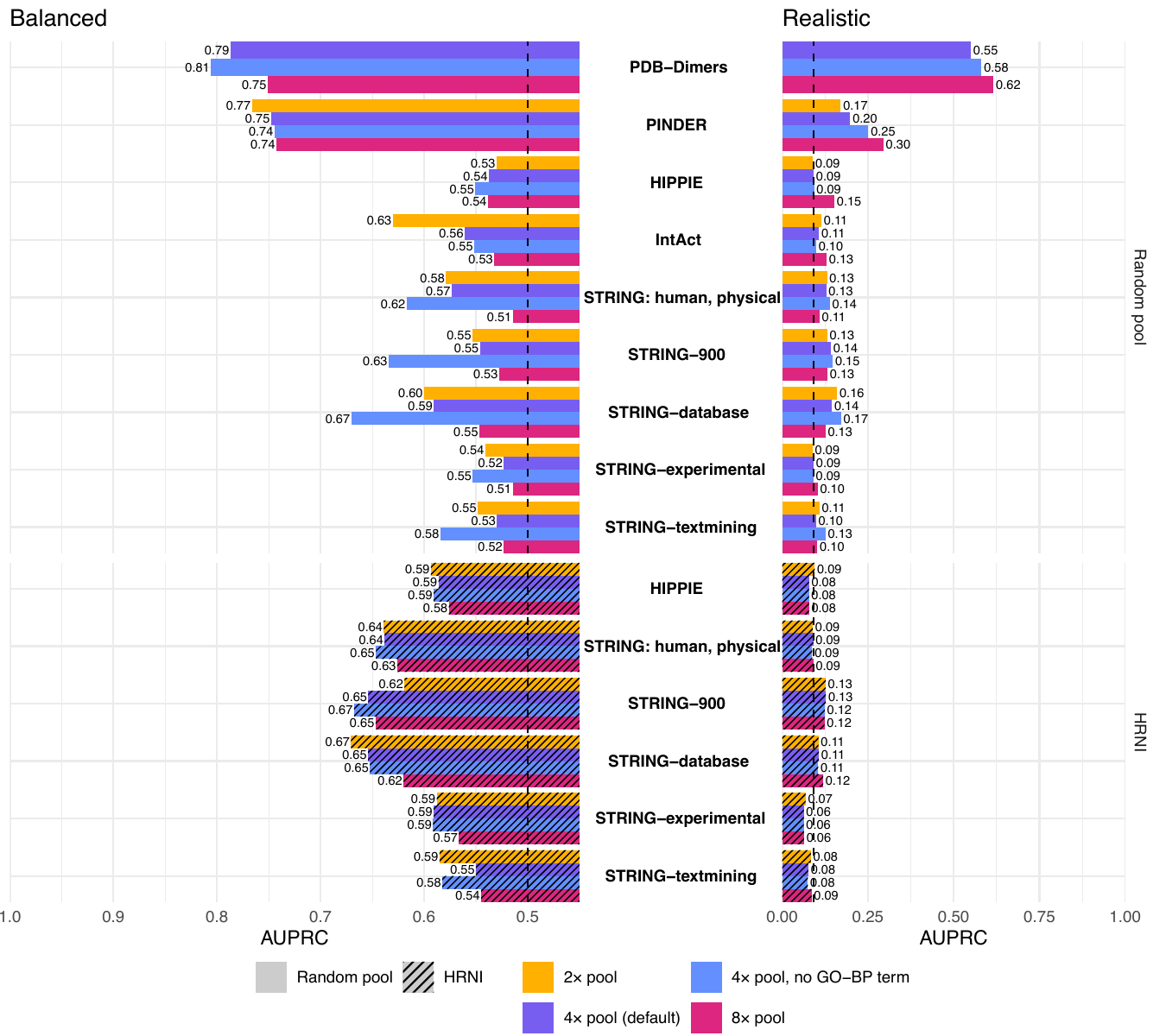}
    \caption{Random Forest classifier results (split/sampling/training seed combination 1234/42/42) comparing the setting in the main text (sampling from a candidate pool $4\timesend$ the number of positives with all bias terms) to settings without the GO-BP term, smaller ($2\timesend$), and larger ($8\timesend$) pool sizes. The exclusion of the GO term mostly inflates the random pool results for the STRING datasets, especially for the ones with the highest shortcut potential for GO-BP overlap, i.\,e., STRING-900, STRING-database, and STRING-textmining (see Figure \ref{fig:ilp_sampling}a). On the corresponding realistic test sets, it can be seen that this responds to little true signal. For the HRNI pool, no considerable changes can be seen since the full HRNI is already lacking overlapping GO-BP terms. For the non-STRING datasets, excluding the GO-BP term does not improve performance considerably, especially on the realistic test set. Hence, including the GO-BP term does not deprive us of too much true signal. Regarding the pool size, sampling from a larger pool size makes it easier to minimize the biases and hence, leads to more unbiased datasets (i.\,e., worse baseline performances).}
    \label{fig:pool_go_rf}
\end{figure}

\begin{figure}[htb]
    \centering
    \includegraphics[width=\linewidth]{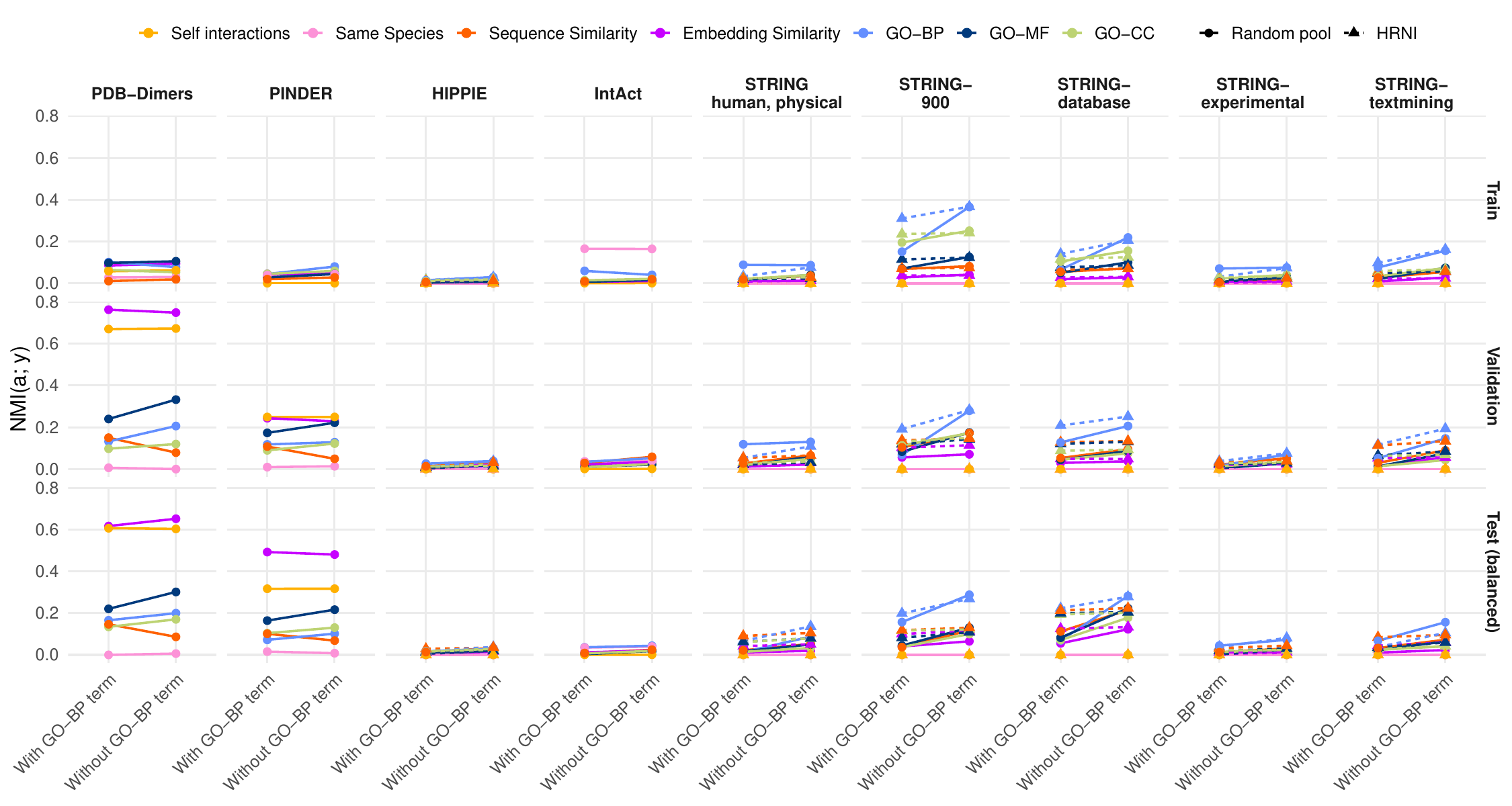}
    \caption{Changes in normalized mutual information when the ILP sampling is run with or without the GO-BP term. Corresponding to the inflated results of Figure \ref{fig:pool_go_rf}, the mutual information for GO-BP (and also GO-CC) goes up fro STRING-900, STRING-database, and STRING-textmining.}
    \label{fig:go_nmi}
\end{figure}

\clearpage

\section{Results for ProtT5 embeddings}\label{appendix-prott5}
\setcounter{figure}{0}
\renewcommand{\thefigure}{B\arabic{figure}}
\setcounter{table}{0}
\renewcommand{\thetable}{B\arabic{table}}

\begin{figure}[!htb]
    \centering
    \includegraphics[width=\linewidth]{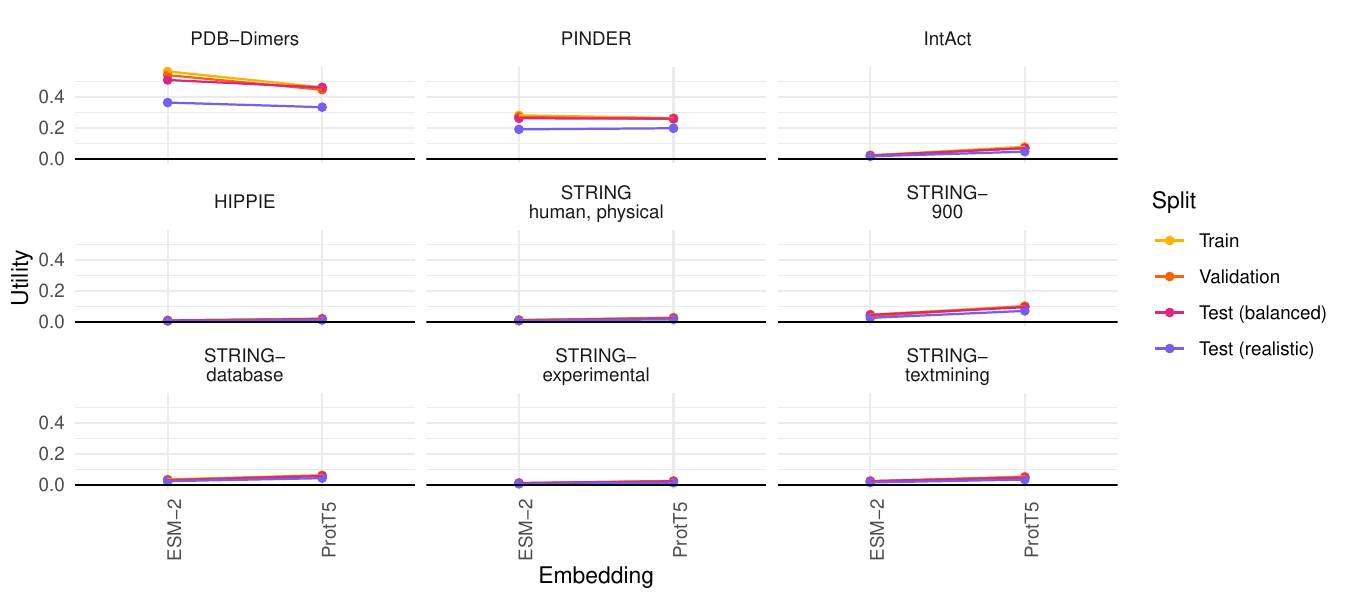}
    \caption{Comparison of the utility of the embedding similarity attribute for ESM-2 versus ProtT5 (split seed: 1234, sampling seed: 42). Overall, the utility is comparable.}
    \label{fig:prott5_nmi}
\end{figure}

\begin{figure}[!htb]
    \centering
    \includegraphics[width=\linewidth]{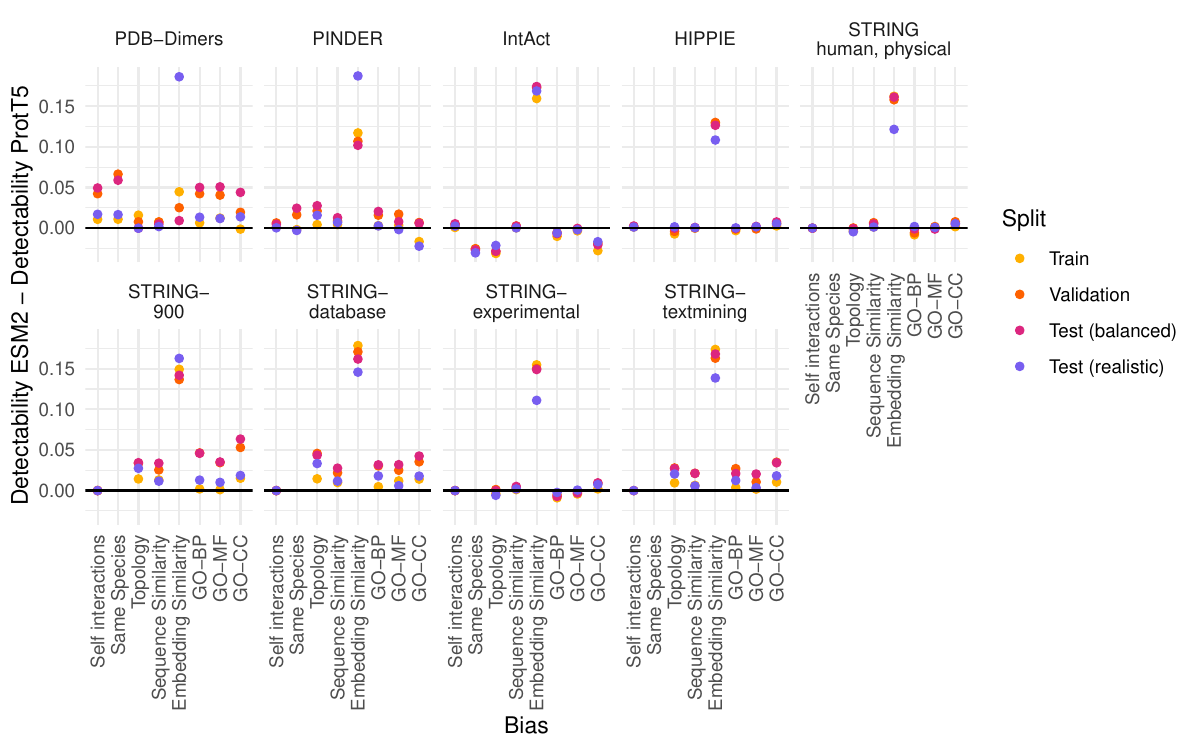}
    \caption{Comparison of the detectability of the different bias attributes in the maximally biased setting. The ridge regressor is either applied from concatenated ESM-2 embeddings or concatenated ProtT5 embeddings (split seed: 1234, sampling seed: 42). Overall, biases are more detectable from ESM-2.}
    \label{fig:prott5_detectability}
\end{figure}

\begin{figure}[!htb]
    \centering
    \includegraphics[width=\linewidth]{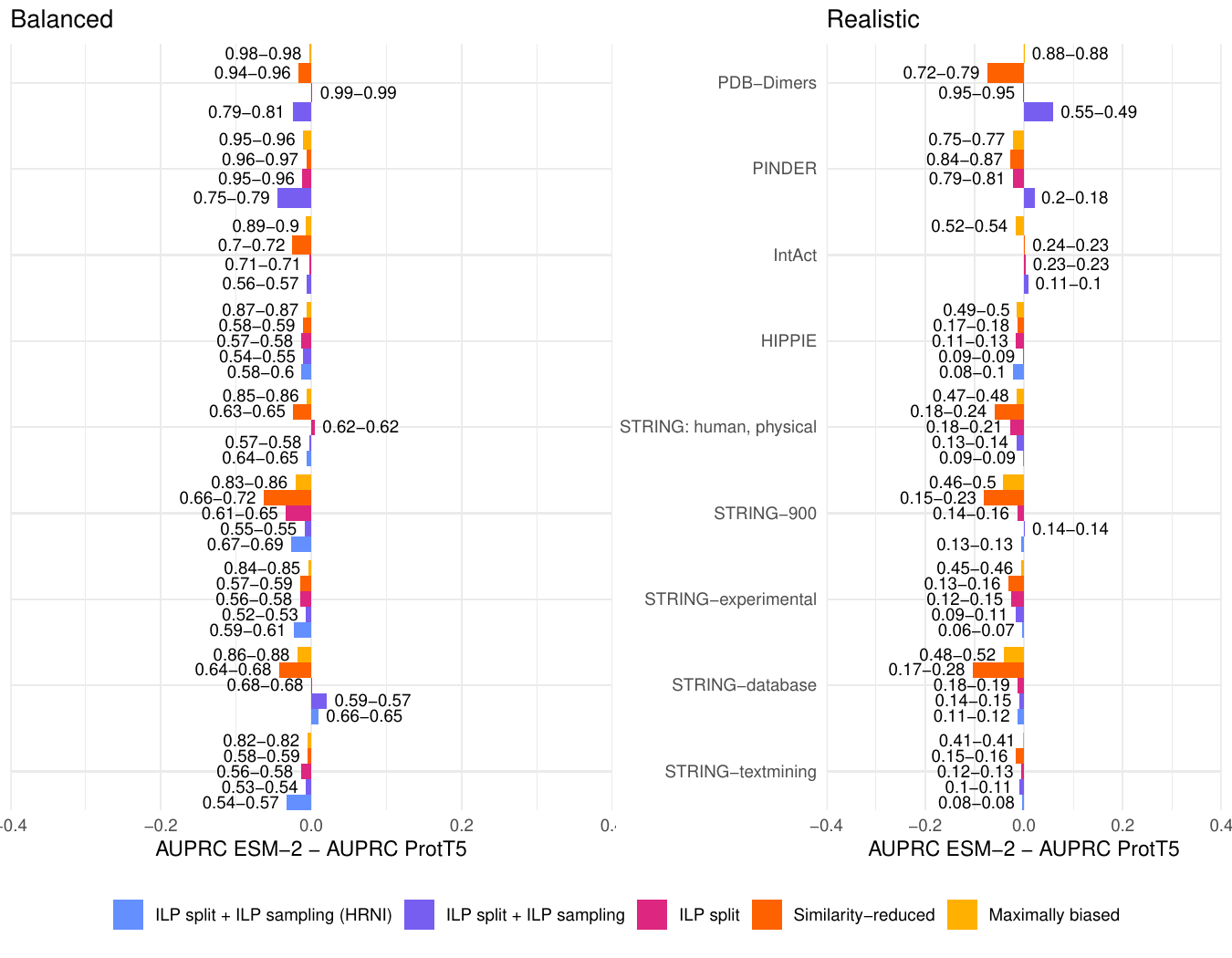}
    \caption{Comparison of the baseline Random Forest classifier results for the ESM-2 versus the ProtT5 embeddings  (split seed: 1234, sampling seed: 42, training seed: 42) . Overall, ProtT5 results are marginally better.}
    \label{fig:prott5_rf}
\end{figure}

\clearpage

{\small\setlength{\tabcolsep}{3pt}

\begin{longtable}[t]{>{\raggedright\arraybackslash}p{2.5cm}llllllll}
\caption{Performance metrics of the baseline with ProtT5 embeddings across datasets, splitting strategies, and test sets on the training, validation, and test datasets generated with the seeds 1234 and 42. The Random Forest seed was 42, too. \textit{(continued on next page)}}\\
\toprule
Category & Test set & AUROC & AUPRC & F1 & MCC & Precision & Recall & Accuracy\\
\midrule
\endfirsthead
\caption[]{Performance metrics of the baseline with ProtT5 embeddings across datasets, splitting strategies, and test sets. \textit{(continued)}}\\
\toprule
Category & Test set & AUROC & AUPRC & F1 & MCC & Precision & Recall & Accuracy\\
\midrule
\endhead

\endfoot
\bottomrule
\endlastfoot
\multicolumn{9}{l}{Dataset: \textbf{PDB-Dimers}}\\
\midrule
 & balanced & 0.98 & 0.98 & 0.94 & 0.88 & 0.95 & 0.94 & 0.94\\

\multirow[t]{-1}{*}[\normalbaselineskip]{\raggedright\arraybackslash \makecell[tl]{Maximally\\biased}} & realistic & 0.98 & 0.88 & 0.78 & 0.76 & 0.66 & 0.94 & 0.95\\

 & balanced & 0.95 & 0.96 & 0.88 & 0.78 & 0.95 & 0.82 & 0.88\\

\multirow[t]{-1}{*}[\normalbaselineskip]{\raggedright\arraybackslash \makecell[tl]{Similarity-\\reduced}} & realistic & 0.94 & 0.79 & 0.65 & 0.63 & 0.54 & 0.82 & 0.92\\

 & balanced & 0.99 & 0.99 & 0.96 & 0.92 & 0.97 & 0.95 & 0.96\\

\multirow[t]{-1}{*}[\normalbaselineskip]{\raggedright\arraybackslash ILP split} & realistic & 0.99 & 0.95 & 0.85 & 0.84 & 0.78 & 0.95 & 0.97\\

 & balanced & 0.80 & 0.81 & 0.76 & 0.46 & 0.70 & 0.84 & 0.73\\

\multirow[t]{-1}{*}[\normalbaselineskip]{\raggedright\arraybackslash \makecell[tl]{ILP split +\\ILP sampling}} & realistic & 0.77 & 0.49 & 0.24 & 0.17 & 0.14 & 0.84 & 0.49\\

\midrule
\multicolumn{9}{l}{Dataset: \textbf{PINDER}}\\
\midrule
 & balanced & 0.96 & 0.96 & 0.90 & 0.81 & 0.91 & 0.90 & 0.90\\

\multirow[t]{-1}{*}[\normalbaselineskip]{\raggedright\arraybackslash \makecell[tl]{Maximally\\biased}} & realistic & 0.96 & 0.77 & 0.63 & 0.61 & 0.48 & 0.90 & 0.90\\

 & balanced & 0.96 & 0.97 & 0.91 & 0.82 & 0.92 & 0.90 & 0.91\\

\multirow[t]{-1}{*}[\normalbaselineskip]{\raggedright\arraybackslash \makecell[tl]{Similarity-\\reduced}} & realistic & 0.96 & 0.87 & 0.66 & 0.65 & 0.52 & 0.90 & 0.92\\

 & balanced & 0.96 & 0.96 & 0.89 & 0.78 & 0.86 & 0.93 & 0.89\\

\multirow[t]{-1}{*}[\normalbaselineskip]{\raggedright\arraybackslash ILP split} & realistic & 0.96 & 0.81 & 0.49 & 0.49 & 0.34 & 0.93 & 0.83\\

 & balanced & 0.78 & 0.79 & 0.70 & 0.30 & 0.59 & 0.87 & 0.63\\

\multirow[t]{-1}{*}[\normalbaselineskip]{\raggedright\arraybackslash \makecell[tl]{ILP split +\\ILP sampling}} & realistic & 0.57 & 0.18 & 0.16 & -0.05 & 0.09 & 0.87 & 0.15\\

\midrule
\multicolumn{9}{l}{Dataset: \textbf{IntAct}}\\
\midrule
 & balanced & 0.89 & 0.90 & 0.84 & 0.61 & 0.77 & 0.92 & 0.80\\

\multirow[t]{-1}{*}[\normalbaselineskip]{\raggedright\arraybackslash \makecell[tl]{Maximally\\biased}} & realistic & 0.89 & 0.54 & 0.39 & 0.37 & 0.25 & 0.92 & 0.69\\

 & balanced & 0.76 & 0.72 & 0.69 & 0.38 & 0.68 & 0.70 & 0.69\\

\multirow[t]{-1}{*}[\normalbaselineskip]{\raggedright\arraybackslash \makecell[tl]{Similarity-\\reduced}} & realistic & 0.75 & 0.23 & 0.30 & 0.23 & 0.19 & 0.70 & 0.67\\

 & balanced & 0.73 & 0.71 & 0.66 & 0.33 & 0.66 & 0.66 & 0.66\\

\multirow[t]{-1}{*}[\normalbaselineskip]{\raggedright\arraybackslash ILP split} & realistic & 0.75 & 0.23 & 0.31 & 0.23 & 0.20 & 0.66 & 0.70\\

 & balanced & 0.56 & 0.57 & 0.64 & 0.07 & 0.51 & 0.83 & 0.52\\

\multirow[t]{-1}{*}[\normalbaselineskip]{\raggedright\arraybackslash \makecell[tl]{ILP split +\\ILP sampling}} & realistic & 0.44 & 0.10 & 0.17 & -0.06 & 0.10 & 0.83 & 0.18\\

\midrule
\multicolumn{9}{l}{Dataset: \textbf{HIPPIE}}\\
\midrule
 & balanced & 0.87 & 0.87 & 0.80 & 0.57 & 0.76 & 0.84 & 0.79\\

\multirow[t]{-1}{*}[\normalbaselineskip]{\raggedright\arraybackslash \makecell[tl]{Maximally\\biased}} & realistic & 0.87 & 0.50 & 0.38 & 0.36 & 0.24 & 0.84 & 0.74\\

 & balanced & 0.60 & 0.59 & 0.48 & 0.14 & 0.60 & 0.39 & 0.57\\

\multirow[t]{-1}{*}[\normalbaselineskip]{\raggedright\arraybackslash \makecell[tl]{Similarity-\\reduced}} & realistic & 0.64 & 0.18 & 0.23 & 0.13 & 0.17 & 0.39 & 0.75\\

 & balanced & 0.58 & 0.58 & 0.41 & 0.11 & 0.59 & 0.32 & 0.55\\

\multirow[t]{-1}{*}[\normalbaselineskip]{\raggedright\arraybackslash ILP split} & realistic & 0.58 & 0.13 & 0.18 & 0.06 & 0.13 & 0.32 & 0.72\\

 & balanced & 0.55 & 0.55 & 0.58 & 0.07 & 0.53 & 0.64 & 0.54\\

\multirow[t]{-1}{*}[\normalbaselineskip]{\raggedright\arraybackslash \makecell[tl]{ILP split +\\ILP sampling}} & realistic & 0.50 & 0.09 & 0.17 & 0.01 & 0.10 & 0.64 & 0.40\\

 & balanced & 0.60 & 0.60 & 0.66 & 0.09 & 0.52 & 0.90 & 0.53\\

\multirow[t]{-1}{*}[\normalbaselineskip]{\raggedright\arraybackslash \makecell[tl]{ILP split +\\sampling (HRNI)}} & realistic & 0.49 & 0.10 & 0.17 & -0.04 & 0.09 & 0.90 & 0.15\\

\midrule
\multicolumn{9}{l}{Dataset: \textbf{STRING: human, physical}}\\
\midrule
 & balanced & 0.85 & 0.86 & 0.78 & 0.55 & 0.78 & 0.78 & 0.78\\

\multirow[t]{-1}{*}[\normalbaselineskip]{\raggedright\arraybackslash \makecell[tl]{Maximally\\biased}} & realistic & 0.85 & 0.48 & 0.38 & 0.35 & 0.26 & 0.78 & 0.77\\

 & balanced & 0.65 & 0.65 & 0.54 & 0.21 & 0.64 & 0.46 & 0.60\\

\multirow[t]{-1}{*}[\normalbaselineskip]{\raggedright\arraybackslash \makecell[tl]{Similarity-\\reduced}} & realistic & 0.68 & 0.24 & 0.26 & 0.17 & 0.18 & 0.46 & 0.76\\

 & balanced & 0.62 & 0.62 & 0.53 & 0.17 & 0.61 & 0.46 & 0.58\\

\multirow[t]{-1}{*}[\normalbaselineskip]{\raggedright\arraybackslash ILP split} & realistic & 0.65 & 0.21 & 0.24 & 0.15 & 0.16 & 0.46 & 0.73\\

 & balanced & 0.57 & 0.58 & 0.53 & 0.10 & 0.55 & 0.50 & 0.55\\

\multirow[t]{-1}{*}[\normalbaselineskip]{\raggedright\arraybackslash \makecell[tl]{ILP split +\\ILP sampling}} & realistic & 0.53 & 0.14 & 0.17 & 0.03 & 0.10 & 0.50 & 0.55\\

 & balanced & 0.66 & 0.65 & 0.68 & 0.20 & 0.55 & 0.88 & 0.58\\*

\multirow[t]{-1}{*}[\normalbaselineskip]{\raggedright\arraybackslash \makecell[tl]{ILP split +\\sampling (HRNI)}} & realistic & 0.45 & 0.09 & 0.15 & -0.10 & 0.08 & 0.88 & 0.12\\

\midrule
\multicolumn{9}{l}{Dataset: \textbf{STRING-900}}\\
\midrule
 & balanced & 0.85 & 0.86 & 0.76 & 0.51 & 0.74 & 0.78 & 0.75\\

\multirow[t]{-1}{*}[\normalbaselineskip]{\raggedright\arraybackslash \makecell[tl]{Maximally\\biased}} & realistic & 0.85 & 0.50 & 0.35 & 0.32 & 0.23 & 0.78 & 0.74\\

 & balanced & 0.73 & 0.72 & 0.68 & 0.34 & 0.66 & 0.71 & 0.67\\

\multirow[t]{-1}{*}[\normalbaselineskip]{\raggedright\arraybackslash \makecell[tl]{Similarity-\\reduced}} & realistic & 0.69 & 0.23 & 0.23 & 0.14 & 0.13 & 0.71 & 0.56\\

 & balanced & 0.66 & 0.65 & 0.62 & 0.23 & 0.61 & 0.63 & 0.61\\

\multirow[t]{-1}{*}[\normalbaselineskip]{\raggedright\arraybackslash ILP split} & realistic & 0.65 & 0.16 & 0.21 & 0.12 & 0.13 & 0.63 & 0.58\\

 & balanced & 0.56 & 0.55 & 0.57 & 0.08 & 0.54 & 0.61 & 0.54\\

\multirow[t]{-1}{*}[\normalbaselineskip]{\raggedright\arraybackslash \makecell[tl]{ILP split +\\ILP sampling}} & realistic & 0.62 & 0.14 & 0.20 & 0.10 & 0.12 & 0.61 & 0.56\\

 & balanced & 0.70 & 0.69 & 0.68 & 0.16 & 0.53 & 0.95 & 0.55\\

\multirow[t]{-1}{*}[\normalbaselineskip]{\raggedright\arraybackslash \makecell[tl]{ILP split +\\sampling (HRNI)}} & realistic & 0.58 & 0.13 & 0.17 & 0.04 & 0.09 & 0.95 & 0.16\\

\midrule
\multicolumn{9}{l}{Dataset: \textbf{STRING-experimental}}\\
\midrule
 & balanced & 0.83 & 0.85 & 0.76 & 0.51 & 0.76 & 0.76 & 0.76\\

\multirow[t]{-1}{*}[\normalbaselineskip]{\raggedright\arraybackslash \makecell[tl]{Maximally\\biased}} & realistic & 0.83 & 0.46 & 0.36 & 0.32 & 0.24 & 0.76 & 0.76\\

 & balanced & 0.58 & 0.59 & 0.49 & 0.11 & 0.57 & 0.42 & 0.55\\

\multirow[t]{-1}{*}[\normalbaselineskip]{\raggedright\arraybackslash \makecell[tl]{Similarity-\\reduced}} & realistic & 0.60 & 0.16 & 0.20 & 0.09 & 0.13 & 0.42 & 0.69\\

 & balanced & 0.58 & 0.58 & 0.52 & 0.10 & 0.56 & 0.48 & 0.55\\

\multirow[t]{-1}{*}[\normalbaselineskip]{\raggedright\arraybackslash ILP split} & realistic & 0.62 & 0.15 & 0.21 & 0.10 & 0.13 & 0.48 & 0.67\\

 & balanced & 0.53 & 0.53 & 0.56 & 0.05 & 0.52 & 0.60 & 0.52\\

\multirow[t]{-1}{*}[\normalbaselineskip]{\raggedright\arraybackslash \makecell[tl]{ILP split +\\ILP sampling}} & realistic & 0.53 & 0.11 & 0.17 & 0.03 & 0.10 & 0.60 & 0.46\\

 & balanced & 0.60 & 0.61 & 0.61 & 0.13 & 0.55 & 0.69 & 0.56\\

\multirow[t]{-1}{*}[\normalbaselineskip]{\raggedright\arraybackslash \makecell[tl]{ILP split +\\sampling (HRNI)}} & realistic & 0.33 & 0.07 & 0.13 & -0.17 & 0.07 & 0.69 & 0.17\\

\midrule
\multicolumn{9}{l}{Dataset: \textbf{STRING-database}}\\
\midrule
 & balanced & 0.88 & 0.88 & 0.80 & 0.59 & 0.76 & 0.85 & 0.79\\

\multirow[t]{-1}{*}[\normalbaselineskip]{\raggedright\arraybackslash \makecell[tl]{Maximally\\biased}} & realistic & 0.88 & 0.52 & 0.38 & 0.36 & 0.24 & 0.85 & 0.74\\

 & balanced & 0.68 & 0.68 & 0.59 & 0.28 & 0.68 & 0.52 & 0.64\\

\multirow[t]{-1}{*}[\normalbaselineskip]{\raggedright\arraybackslash \makecell[tl]{Similarity-\\reduced}} & realistic & 0.70 & 0.28 & 0.29 & 0.21 & 0.20 & 0.52 & 0.76\\

 & balanced & 0.69 & 0.68 & 0.50 & 0.27 & 0.73 & 0.38 & 0.62\\

\multirow[t]{-1}{*}[\normalbaselineskip]{\raggedright\arraybackslash ILP split} & realistic & 0.68 & 0.19 & 0.26 & 0.17 & 0.19 & 0.38 & 0.80\\

 & balanced & 0.58 & 0.57 & 0.46 & 0.10 & 0.57 & 0.38 & 0.55\\

\multirow[t]{-1}{*}[\normalbaselineskip]{\raggedright\arraybackslash \makecell[tl]{ILP split +\\ILP sampling}} & realistic & 0.63 & 0.15 & 0.22 & 0.11 & 0.15 & 0.38 & 0.75\\

 & balanced & 0.65 & 0.65 & 0.65 & 0.20 & 0.57 & 0.75 & 0.59\\

\multirow[t]{-1}{*}[\normalbaselineskip]{\raggedright\arraybackslash \makecell[tl]{ILP split +\\sampling (HRNI)}} & realistic & 0.53 & 0.12 & 0.16 & -0.00 & 0.09 & 0.75 & 0.29\\

\midrule
\multicolumn{9}{l}{Dataset: \textbf{STRING-textmining}}\\
\midrule
 & balanced & 0.81 & 0.82 & 0.74 & 0.47 & 0.73 & 0.75 & 0.74\\

\multirow[t]{-1}{*}[\normalbaselineskip]{\raggedright\arraybackslash \makecell[tl]{Maximally\\biased}} & realistic & 0.81 & 0.41 & 0.34 & 0.30 & 0.22 & 0.75 & 0.73\\

 & balanced & 0.59 & 0.59 & 0.54 & 0.13 & 0.57 & 0.51 & 0.56\\

\multirow[t]{-1}{*}[\normalbaselineskip]{\raggedright\arraybackslash \makecell[tl]{Similarity-\\reduced}} & realistic & 0.63 & 0.16 & 0.21 & 0.10 & 0.13 & 0.51 & 0.65\\

 & balanced & 0.57 & 0.58 & 0.52 & 0.10 & 0.56 & 0.49 & 0.55\\

\multirow[t]{-1}{*}[\normalbaselineskip]{\raggedright\arraybackslash ILP split} & realistic & 0.58 & 0.13 & 0.18 & 0.06 & 0.11 & 0.49 & 0.61\\

 & balanced & 0.55 & 0.54 & 0.57 & 0.07 & 0.53 & 0.63 & 0.54\\

\multirow[t]{-1}{*}[\normalbaselineskip]{\raggedright\arraybackslash \makecell[tl]{ILP split +\\ILP sampling}} & realistic & 0.56 & 0.11 & 0.18 & 0.06 & 0.11 & 0.63 & 0.49\\

 & balanced & 0.57 & 0.57 & 0.62 & 0.09 & 0.53 & 0.75 & 0.54\\

\multirow[t]{-1}{*}[\normalbaselineskip]{\raggedright\arraybackslash \makecell[tl]{ILP split +\\sampling (HRNI)}} & realistic & 0.43 & 0.08 & 0.15 & -0.06 & 0.08 & 0.75 & 0.22
\label{tab:prott5_performance}
\end{longtable}}

\section{Runtime and resource consumption} \label{appendix-runtime}
\setcounter{figure}{0}
\renewcommand{\thefigure}{C\arabic{figure}}
\setcounter{table}{0}
\renewcommand{\thetable}{C\arabic{table}}

The ILP configurations are more expensive than their alternatives. Figure~\ref{fig:runtime_resources} reports wall-clock runtime and peak memory per pipeline step across the five settings (maximally biased, similarity-reduced, ILP split, ILP split + ILP sampling from the complement, and ILP split + ILP sampling from the HRNI pool). 

\begin{figure}[!htb]
    \centering
    \includegraphics[width=\linewidth]{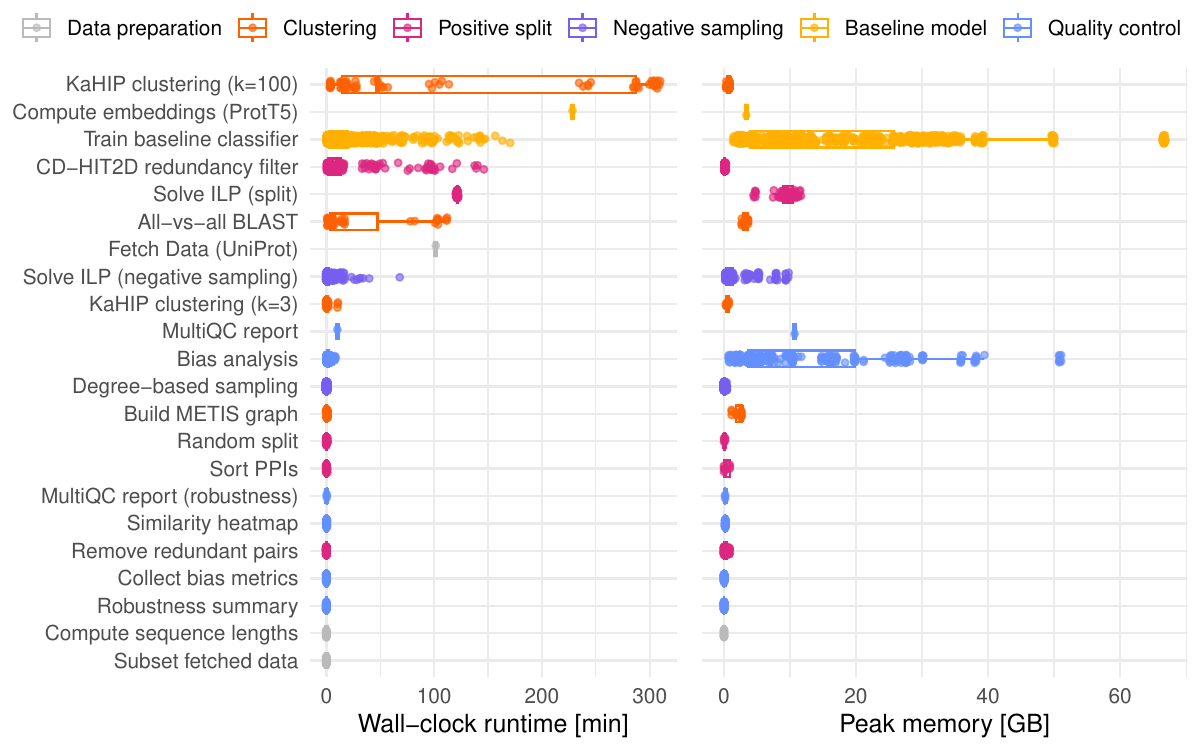}
    \caption{Runtime and resources taken up by the individual pipeline steps for all configurations reported in the main text. The numbers come from the Nextflow execution reports of the ESM-2 robustness runs: 9 datasets x 4 splitting strategies, plus ILP split + sampling from the HRNI network for 6 datasets, each with 3 split seeds x 3 sampling seeds x 3 training seeds, and the corresponding bias analyses on the single split/sampling seed combination. Two kinds of steps do not occur in the robustness runs and come from the ProtT5 pipeline run on all datasets: computing protein embeddings (a single-task embedding of all proteins; the ESM-2 embeddings were precomputed) and generating quality-control reports (similarity heatmap, collecting bias metrics, MultiQC report). }
    \label{fig:runtime_resources}
\end{figure}

The embedding computation, solving the ILP, fetching the data, and the KaHIP clustering into 100 clusters (which is only required for the ILP split) have the longest median runtime. 
Solving the ILP for the positive split always reached the 2-hour time limit, so the reported cluster assignments are feasible but not optimal. 
The negative-sampling ILP is fast on the small validation and test sets but takes up to one hour, depending on the dataset size and bias terms. This is why we cap the pool at four times the positive count (here, up to $2$ million for the training set), rather than drawing from the full complement or the full HRNI set. The pool is subsampled in a way that makes it possible to find a feasible solution to the ILP. Depending on the activated bias terms, negative self-interactions are added, pairs with a nonzero GO-BP Jaccard index are preferred, and taxa and positive degrees are taken into account (for more details, see Section \ref{sec:sample-neg}). The influence of the pool size on the solution can be seen in Figures \ref{fig:pool_size_obj} and \ref{fig:pool_go_rf}. For the largest pool size ($3.5$ million $=8 \times$ the positive count of IntAct), the ILP ran for $81$ minutes. However, on HIPPIE, which has a similar pool size, the ILP finished in 10 minutes, showing the considerable effect of the added taxonomy bias term for the IntAct ILP. 
The embeddings are generated once for all sequences combined (all datasets: $\num{259687}$ sequences). The runtime of training the baseline classifier scales with the number of interactions, whereas all-vs-all BLAST and the CD-HIT-2D redundancy filter scale with the number of proteins, as every protein is compared with every other.  

Memory consumption is highest for baseline training, which holds the concatenated embeddings of all pairs in memory, followed by the bias analysis. Both are highest for HIPPIE, IntAct, STRING, and STRING-experimental the largest datasets. All other steps stay below 15~GB.
All ILPs were solved with Gurobi under an academic license. CVXPY supports open alternatives, but these are typically slower on problems of this size, so users without a commercial solver should expect longer runtimes than those reported here. All analyses ran on a heterogeneous SLURM cluster comprising Intel Xeon Gold 6148 (2.40\,GHz), AMD EPYC 7351, and AMD EPYC 7713 nodes, and on a single node with two AMD EPYC 7402 processors (24 cores each, up to 3.35\,GHz) and 503\,GB RAM, of which 26 hardware threads were used, all using conda environments. Embedding generation used an NVIDIA A40 GPU. The main text results took about 770 task hours and 4,300 CPU hours to complete. 

\end{document}